# Structural stability, elemental ordering, and transport properties of layered $ScTaN_2$

Baptiste Julien*, Ian A. Leahy, Rebecca W. Smaha, John S. Mangum, Craig L. Perkins, Sage R. Bauers, Andriy Zakutayev*

*Materials Science Center, National Renewable Energy Laboratory, 15013 Denver West Parkway, Golden, Colorado 80401, United States.*

**ABSTRACT**. Ternary transition metal (TM) nitrides have gained significant attention in thin film research due to their promising properties for a broad range of applications. Particularly, some of the ternary TM nitrides are predicted to adopt layered structures that make them interesting for thermoelectric conversion and quantum materials applications. Unfortunately, synthesis of TM ternary nitride films by physical vapor deposition often favors disordered 3D structures rather than the predicted 2D-like layered structure. In this study, we investigate the structural interplay in the Sc-Ta-N material system, focusing on the $ScTaN_2$ composition. We use a two-step approach to prepare Sc-Ta-N materials: first deposit combinatorial films by radio-frequency co-sputtering and then process the resulting 3D-structured samples with rapid thermal annealing (RTA). Synchrotron grazing-incidence wide-angle X-ray scattering (GIWAXS) on films annealed at 1200 °C for 20 min reveals the nucleation of the layered structure ($P6_3/mmc$) within a composition window of $x$ = Sc/(Sc+Ta) = 0.2 – 0.5. We estimate the long-range order parameter in stoichiometric $ScTaN_2$ to be 0.86, corresponding to a fraction of Sc/Ta antisites of 7%. Interestingly, we find that the structure can accommodate large off-stoichiometry in the Ta-rich region ($x < 0.5$), facilitated by making an alloy with the nearly isostructural $Ta_5N_6$ compound that exists on a composition tie-line at $x$ = 0. Transport measurements on $ScTaN_2$ reveal a nearly temperature-independent high carrier density ($10^{21}$ $cm^{-3}$), suggesting a heavily doped semiconductor or semimetallic character. The carrier mobility is relatively small (9.5 $cm^2V^{-1}s^{-1}$ at 2 K) and the residual-resistivity ratio is small, suggesting that electrical conduction is dominated by defects or disorder. Measured magnetoresistance is indicative of weak antilocalization at low temperatures. This paper highlights the interplay between $ScTaN_2$ and $Ta_5N_6$ crystal structures in stabilizing layered structures, emphasizes the importance of cation order/disorder for potential tunable alloys, and suggests that $ScTaN_2$ is promising platform for exploring electronic properties in a tie line of stoichiometry.

## I. INTRODUCTION

Ternary nitrides combining two metal cations with a nitrogen anion are promising materials due to their unique electronic and thermal properties, involving a broad range application such as ceramic coatings [1], [2], semiconductors [3], [4], permanent magnets [5], [6], spintronics [7], topological materials [8], superconductors [9], [10], [11], and more. Many of ternary nitrides adopt similar structures to well-known layered oxides, widely investigated for energy applications [12]. For example, $ABN_2$ nitrides (where $B$ is a transition metal) tend to adopt layered structures comprised of alternating sheets of edge-sharing $AN_6$ octahedra and $BN_6$ trigonal prisms, sometimes referred as “rockseline” (a portmanteau of rocksalt and nickeline), as in the recently synthesized $MgMoN_2$ and $MgWN_2$ thin films [13,14]. This structure is apparent to a 211-MAX phase, such as $Ti_2AlN$ (both being $P6_3/mmc$), where Ti and N occupy inverted Wyckoff positions [15]. Layered structures often exhibit anisotropy between phonon and electron transport, which usually enhance thermoelectric properties, as predicted in layered structures $CuTaN_2$ (delafossite-type) and $NaTaN_2$ (α-$NaFeO_2$-type) [16]. Bulk synthesis and characterization of $CuTaN_2$ have also demonstrated strong optical absorption [17], and confirm overall that layered ternary nitrides are strong candidates for energy harvesting applications.

Despite their promising properties, thin film synthesis of ternary layered nitrides presents a significant challenge. Traditional physical vapor deposition techniques such as reactive sputtering often tend to favor formation of metastable phases with a high degree of cation disorder rather than ordered layered phases [18]. This tolerance to cation disorder in ternary nitrides has already been investigated using statistical polymorph sampler approaches [19], [20]. In particular, it has been demonstrated that in the case of sputtered $ZnZrN_2$, the configurational entropy contributes to the stabilization of a metastable cation-disordered rocksalt structure rather than the predicted stable layered structure [21]. Moreover, first principles calculations and band structure arguments have shown that in the case of metal oxides with octahedrally-coordinated cations, $d^0$ transition metals (such as $Zr^{4+}$, $Ta^{5+}$ or $Mo^{6+}$) can tolerate site distortions with a low energy cost and therefore can stabilize cation-disordered phases [22]. Despite a long-range disorder typically observed experimentally, ternary nitride precursors may have hidden short-range ordering that can be promoted to long-range ordering upon rapid thermal annealing. This has been recently investigated, experimentally and theoretically, for layered $MgMoN_2$. The transition temperature at which the film precursor converts to the stable layered polymorph was found to be highly dependent on the material [13]. Nevertheless, none of those previous thin film studies focus on detailed structural analysis, such as the degree of elemental ordering, phase competition, morphology or electronic properties, which are concepts that are necessary to understand to promote these novel materials into applications.

For the Sc-Ta-N ternary system, the focus of this study, $ScTaN_2$ is the only reported stable ternary phase. $ScTaN_2$ crystallizes in the $P6_3/mmc$ space group (no. 194) and consists of $ScN_6$ octahedra and $TaN_6$ trigonal prismatic sheets, suggesting an ionic picture: $(Sc^{3+})_{oct}(Ta^{3+})_{tri}N_2$. While $ScTaN_2$ has been experimentally synthesized in bulk [23], [24], thin film synthesis has not been investigated yet. Previous calculations predicted a narrow-bandgap semiconductor with a calculated bandgap of 0.139 eV, which is lower than ScN (0.9 eV [25]), as well as an anisotropic Seebeck coefficient with a high theoretical absolute value at room temperature (~200 μV/K). There is an experimentally observed semimetallic behavior of the resistivity, consistent with their observation of antisite defects in their structure refinement [23]. These preliminary results show that $ScTaN_2$ could be a strong candidate for thermoelectric applications.

In this work, we explore the Sc-Ta-N space in sputtered thin film form with help of X-ray scattering analysis and transport property measurements. Using combinatorial sputtering, we synthesize compositionally graded $Sc_xTa_{1-x}N$ films, where $x$ = Sc/(Sc+Ta), subsequently referred to as $ScTaN_2$. We then employ rapid thermal annealing (RTA) in order to promote the formation of the layered structure after 20 min at 1200 °C. Synchrotron grazing-incidence wide-angle X-ray scattering (GIWAXS) reveals an interesting off-stoichiometry stabilization of the layered structure in the Ta-rich region, eventually leading to the formation of quasi-isostructural $Ta_5N_6$. By analyzing the superlattice peak (002), we estimate the long-range order parameter in stoichiometric $ScTaN_2$ to be 0.86, corresponding to a relatively small fraction of antisites (7%). Finally, electrical and magneto-transport reveal $ScTaN_2$ to be lying between a semimetal and a narrow bandgap semiconductor with a hole-type conduction, likely induced by structural defects and impurities. Overall, this work provides insight into the cation ordering in layered nitride thin films and highlights the interplay between $ScTaN_2$ and quasi-isostructural $Ta_5N_6$.

## II. METHODS

### A. Synthesis

$ScTaN_2$ thin film libraries were synthetized by combinatorial radio-frequency (RF) sputtering. 50.8 mm diameter Sc and Ta targets (99.99% pure, Kurt J. Lesker) were 180° opposed from each other and sputtered using RF powers varying between 30 W and 70 W in order to target a specific range of composition. Films around stoichiometric ($x$ = 0.5) were sputtered using a power of 70 W for Sc (corresponding to a power density of 1.48 $Wcm^{-2}$) and 30 W for Ta (corresponding to a power density of 3.45 $Wcm^{-2}$). Ta-rich compositions were targeted by using a target power of 60 W (2.96 $Wcm^{-2}$) for both Sc and Ta. The base pressure in the chamber before deposition was recorded at $1.10^{-7}$ Torr. Ar and $N_2$ gases were introduced to the chamber with flows of 6 and 3 sccm, respectively. The total pressure during deposition was set to 6 mTorr. A RF nitrogen plasma source operating at 350 W was used for enhancing the chemical potential of nitrogen during deposition. The substrates were kept stationary without external heating. Combinatorial libraries were deposited on 50.8 mm × 50.8 mm square Si(100) substrates with an existing 100 nm layer of insulating $SiN_x$. Prior to deposition, the substrates were sequentially sonicated in acetone and isopropyl alcohol for 10 min and rinsed with deionized water. At the end of the deposition, the Sc-Ta-N libraries were capped with a thin amorphous $SiN_x$ layer to prevent reaction with air. During this step, a Si target was sputtered at a power of 60 W and the gas flows were kept the same. The sample was rotated to ensure good homogeneity of the capping layer. Finally, the 2-inch square libraries were cleaved in four equivalent rows of 50.8 mm × 12.7 mm, in the composition gradient direction.

Rapid thermal annealing of the as-deposited Sc-Ta-N libraries was performed using a mini lamp annealer (Advance Riko, MILA-5000) in a flowing $N_2$ atmosphere, with a $N_2$ flow of 10 slpm. The films were first heated at 100 °C for 3 min to evaporate surface contaminants, and then rapidly heated up to the desired annealing temperature with a heating rate of 30-35 °C/s. The dwell time was set to 20 min. At the end, the heating source was automatically turned off and the films were naturally cooled down to room temperature in flowing $N_2$ atmosphere. A schematic of the RTA setup is presented in Fig. S1. Such annealing resulted in the vaporization of the thin capping layer.

### B. Characterization

The cation composition was characterized by X-ray fluorescence (XRF) using a Fischer XDV-SDD operating with a Rh source. The Sc/Ta concentration was measured at equally spaced points along the compositionally graded sample. Elemental spectra were acquired with a spot size of 3 mm and an exposure time of 120 s. We define the cation fraction $x$ = Sc/(Sc+Ta), so that $x$ = 0.5 corresponds to stochiometric $ScTaN_2$. The anionic ratio O/(O+N) in the film was quantified by depth-profiling Auger electron spectroscopy (AES) on a Physical Electronics 710 system operating with a 10 kV and 10 nA primary electron beam.

Structural characterizations of combinatorial libraries were primarily performed by laboratory X-ray diffraction (XRD) with a Bruker D8 diffractometer equipped with an area detector and a Cu Kα radiation source (λ = 1.5406 Å). The samples were mounted on a mapping stage and 2D diffraction images were sequentially acquired across the libraries in the θ-2θ geometry, with an exposure of 150 s for each point. The detector images were then integrated in the azimuthal angle (χ) to produce diffraction patterns. Libraries of interest were further characterized by grazing-incidence wide-angle X-ray scattering (GIWAXS) at the Stanford Synchrotron Radiation Lightsource (SSRL, beamline 11-3), using a 12.700 keV radiation source (λ = 0.97625 Å) and a Rayonix MX225 CCD area detector. The data were collected with a 1° incident angle, a sample-to-detector distance of 150 mm, and a spot size of 50 μm by 150 μm. Synchrotron GIWAXS data were processed with the *Nika* and *Irena* packages [26,27], and combinatorial data sets were analyzed and processed using the high-throughput data analysis package *COMBIgor* [28]. LeBail refinements were performed in *GSAS-II* [29].

$ScTaN_2$ layered structures with different degrees of ordering were simulated using a custom-built Python script, by varying the occupancy of the four different cation sites: *oct*-Sc, *tri*-Ta, *oct*-Ta and *tri*-Sc. The overall stoichiometry was kept to $ScTaN_2$. For each structure, we calculated the powder diffraction pattern (PXRD) and extracted the theoretical peak intensities. PXRD patterns were calculated using the Python library *pymatgen*. The isotropic thermal parameter B was set to 1 for each atom and the radiation wavelength to λ = 0.97625 Å, for comparison with synchrotron data.

Cross-sectional observation of selected samples was performed by scanning electron microscopy (SEM), on a ThermoFisher Nova 630, operating at 5 kV with a current of 0.45 nA. The films were first mechanically cleaved and then polished by ion beam milling on a JEOL cross-section polisher. Energy dispersive spectroscopy (EDS) maps of cross-sections were performed with an Oxford Ultim Max EDS detector operating at 15 kV accelerated voltage.

Electrical and magneto transport measurements were performed in a Quantum Design Physical Property Measurement System (PPMS) in the 2 – 300 K temperature range. A near-stoichiometry film of $ScTaN_2$ film (~140 nm thick) was contacted with a Van der Pauw geometry. The magnetic field was swept from -14 T to 14 T. The carrier density and mobility were extracted from the Hall resistance and measured as a function of temperature. Seebeck coefficient was measured at room temperature using a custom set up.

## III. RESULTS AND DISCUSSION

### A. Synthesis of $ScTaN_2$ film precursors

Combinatorial sputtering on Si/$SiN_x$ substrates at room temperature yielded $ScTaN_2$ films with composition ranging from $x$ = Sc/(Sc+Ta) = 0.35 to $x$ = 0.71, as characterized by XRF. Laboratory XRD patterns across a combinatorial film are shown in Fig. 1(a). For clarity, the powder diffraction pattern (PXRD) of a disordered rocksalt $ScTaN_2$ (referred as RS), with a lattice parameter based on bulk ScN (space group *Fm*-3*m*), is displayed below the experimental data. The as-deposited films exhibit a polycrystalline rocksalt structure over the entire range of compositions. In fact, ScN and TaN adopt rocksalt structures separately, so a rocksalt structure would also likely be observed in a cation-disordered alloy. On the other hand, cation ordering in the ternary phase induces structural distortions from cubic rocksalt that would lead to additional reflections. Thus, the average structure probably consists of a disordered rocksalt structure (referred to as RS-$ScTaN_2$), as displayed in Fig. 1(c), in which Sc and Ta randomly occupy the N-coordinated octahedra. This observation is consistent with previous work on sputtered ternary nitrides [18]. As the Sc fraction increases in the film, the (111) and (200) peaks shift to lower angle, suggesting a lattice expansion. The lattice constant of the cubic structure is shown in Fig. 1(b). It increases linearly, following a Vegard-like behavior, confirming that the as-deposited film can be seen as a solid solution with a rocksalt structure where Sc and Ta are randomly distributed.

The changes in intensity of (111) and (200) peaks suggest that the film gets preferentially oriented along (200) as the Sc relative concentration increases.

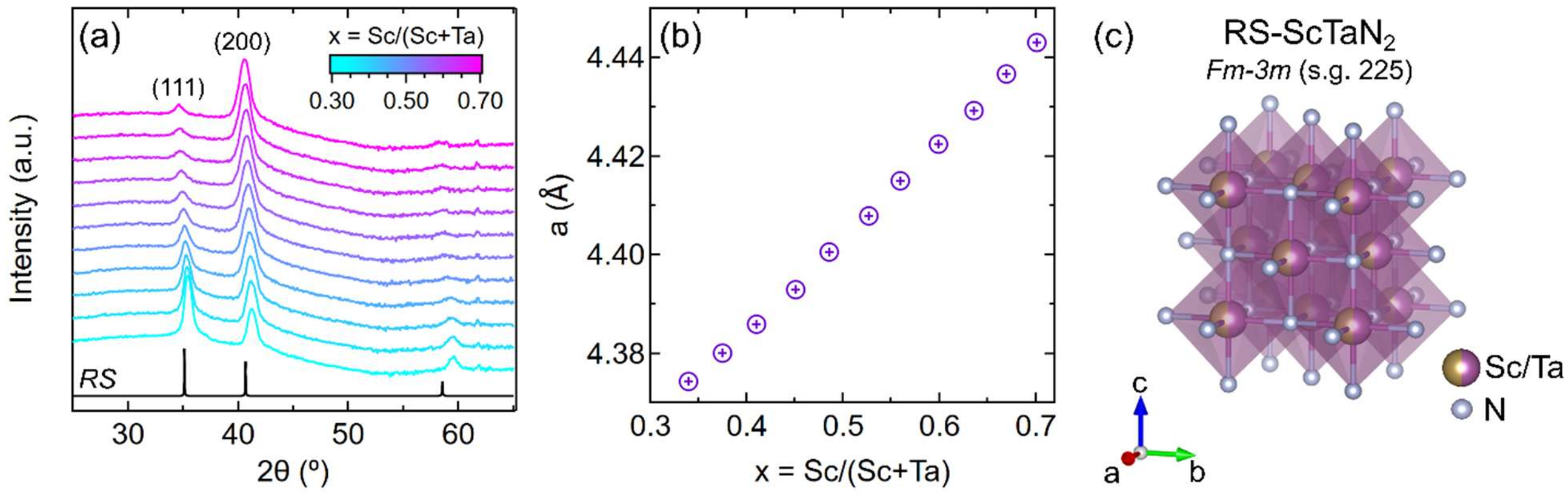

FIG. 1. (a) Laboratory XRD of as-deposited combinatorial $ScTaN_2$ films. The reference rocksalt pattern is taken from bulk ScN. (b) Evolution of the rocksalt lattice parameter with the Sc fraction. (c) Crystal representation of disordered rocksalt $ScTaN_2$, referred to as RS-$ScTaN_2$.

### B. Annealing and structural characterization of layered $ScTaN_2$ films

The $ScTaN_2$ combinatorial films were then processed by RTA under flowing $N_2$ at different temperatures (1000 °C, 1100 °C, and 1200 °C), with a dwell time of 20 min. For each temperature, the different phases were identified by laboratory XRD, across the film composition $x$, and then reported on the phase map shown in Fig. 2(a). The acronyms RL refers to the layered phase (so-called "rockseline" $P6_3/mmc$). Up to 1100 °C, the $ScTaN_2$ films remain in a rocksalt structure. As shown in Fig. S3, the lattice parameter of annealed films is smaller than the as-deposited films. This effect could be due to stress relaxation or the reduction of native defects and impurities upon annealing. The phase transformation occurs at 1200 °C, where the RS phase converts to a phase-pure RL in the Ta-rich composition region, from $x = 0.35$ to $x = 0.50$. Beyond that point, only a mixture of RS and RL or phase-pure RS is observed. Although a complete time-dependance study is beyond the scope of this work, additional annealing was performed at 1200 °C for a limited time of 3 min and resulted in an incomplete phase transformation, as shown in Fig. S4. Synchrotron GIWAXS data of films annealed at 1200 °C is presented in Fig. 2(b). The experimental patterns for $x < 0.57$ match with the RL phase, as illustrated by the three fundamental reflections (101), (102) and (103) located at $q = 2.46$ Å$^{-1}$, 2.68 Å$^{-1}$, and 2.99 Å$^{-1}$, respectively. Furthermore, the presence of the low-angle superlattice peak (002) at $q = 1.2$ Å$^{-1}$ suggests the long-range ordering of Sc and Ta in their respective layers.

For composition-dependent analysis, the intensities of the characteristic peaks (002) and (004), referred to as RL(002) and RL(004), and the rocksalt peak (200), referred to as RS(200), were extracted, normalized, and plotted against composition in Fig. 2(c). For $x > 0.57$, the material remains phase-pure RS. The RS → RL transition occurs around $x = 0.57$, as RL(002) and RL(004) peaks are emerging. At the same time the RS(200) intensity starts to drop, and the coexistence of RS and RL phases is observed for $0.54 \leq x \leq 0.57$. The RS phase vanishes completely at $x = 0.50$ and phase-pure RL is observed in the Ta-rich region. If we look at the RL(002) intensity, the optimal condition, supposedly the highest degree of ordering, is found to be $0.42 < x$ 0.50, with a maximum at $x = 0.46$, and thus not exactly centered at $ScTaN_2$. This could be indicative of a slight nitrogen deficiency in the structure. As shown in Fig. S5, a certain degree of texture is present. Because of this texture effect as well as the missing wedge in GIWAXS experiment, the intensity of (10$l$) peaks relative to (002) and (004) are underestimated. For $x \leq 0.38$, we note the emergence of two weak reflections (*) on Fig. 2(b), located at $q = 1.41$ Å$^{-1}$ and $q = 1.85$ Å$^{-1}$. These two reflections can be indexed to binary $Ta_5N_6$ (space group $P6_3/mcm$), which is quasi-isostructural to $ScTaN_2$ and discussed more in later sections.

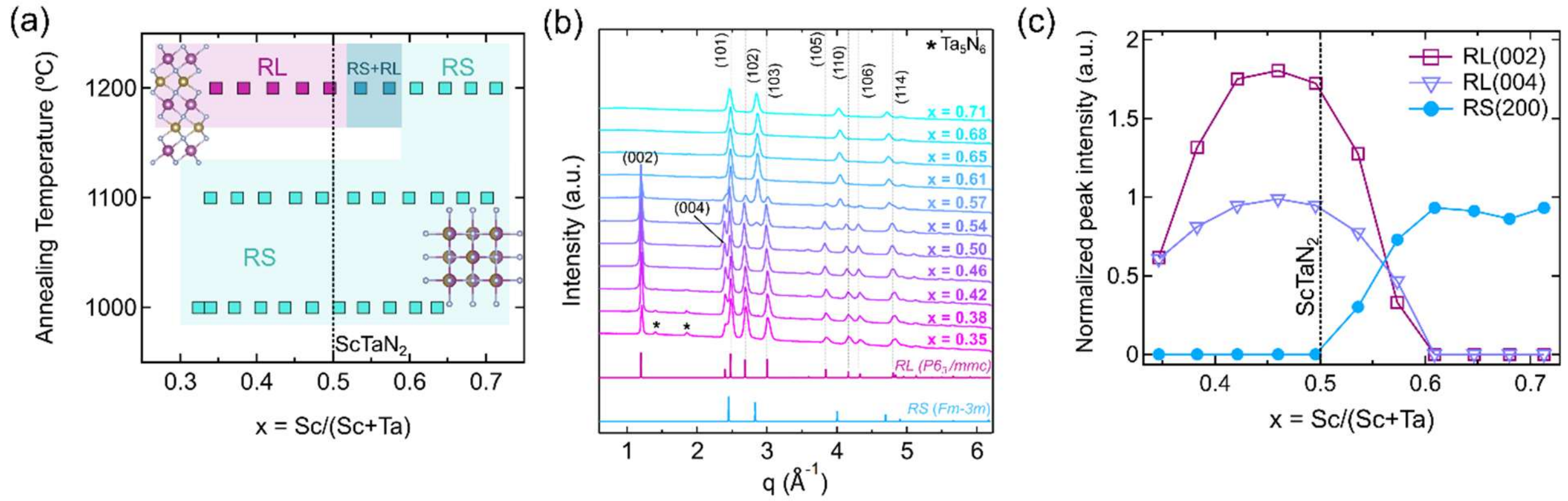

FIG. 2. Structure characterization of $ScTaN_2$ combinatorial films after RTA. (a) Structural phase map with respect to the annealing temperature and composition. (b) Synchrotron integrated GIWAXS patterns on films annealed at 1200 °C (λ = 0.97625 Å). (c) XRD peak analysis on the film annealed at 1200 °C, showing RL(002), RL(004) and RS(200) peak intensity, with RS and RL referring to the rocksalt and the "rockseline" phase respectively. For accurate comparison, the intensity of each peak is normalized by [RL(101) + RS(111)].

We refined the lattice parameters by performing a LeBail fit in the $ScTaN_2$ RL structure at the stoichiometric condition ($x$ = 0.5), because full Rietveld refinement here is not accurate due to the deviation in peak intensities caused by preferential crystallographic orientation. The result of the LeBail refinement is shown in Fig. 3(a); it yielded $a$ = 3.0151(4) Å and $c$ = 10.4710(4) Å with $R_{wp}$ = 4.041%. The volume of the unit cell was $V$ = 82.436(2) Å$^3$. This is slightly smaller than the one previously reported for bulk synthesis of $ScTaN_2$ by Niewa *et al.* ($V$ = 85.33 Å$^3$) [23], possibly due to stress relief or recrystallisation induced during RTA.

Although the appearance of the superlattice peak (002) in the annealed films is evidence of long-range order, the degree of ordering is likely to be lower than the theoretical structure, due to anomalies such as atomic sites being occupied by the "wrong" atoms [30]. Focusing on stoichiometric $ScTaN_2$, we introduce the quantity δ in $(Sc_{1-\delta}Ta_{\delta})_{oct}(Ta_{1-\delta}Sc_{\delta})_{tri}N_2$, defined as the fraction of "swapped sites", or so-called antisites, i.e. Ta located in an octahedron normally occupied by Sc, and vice versa, in equal concentration fixed by $ScTaN_2$ stoichiometry. The subscripts *oct* and *tri* refer to the octahedral and trigonal prismatic sites. δ = 0 corresponds to the complete ordered structure whereas δ = 0.5 corresponds to complete disorder, with Sc and Ta randomly located in either site in equal proportion. The overall composition was fixed at ideal stoichiometry (i.e. $x$ = 0.5) and the nitrogen to cation ratio is assumed to be N/(Sc+Ta) = 1.

The degree of long-range order can be estimated by looking at the relative intensities of the (00*l*) reflections. In $ScTaN_2$ layered structure, where Sc and Ta are ordered in their respective layers, the (002) superlattice peak corresponds to the *d*-spacing of each identical layer ($d_{002}$ = $c$/2 = 5.29 Å) and is indicative of separation of the Sc and Ta cations within their respective layer. In contrast, the (004) peak corresponds to the *d*-spacing of each adjacent layer ($d_{004}$ = $c$/4 = 2.64 Å) and is only indicative of the layered nature of the structure regardless of Sc/Ta site occupancy. Hence, while the (002) reflection is strongly influenced by the degree of cation ordering, the (004) peak is not sensitive to it, with respect to atomic structure factors. Comparing (002) and (004) intensities is also not sensitive to the texture effect along [00*l*] direction. Thus, we define the long-range order parameter $S$ as:

$$S = \sqrt{\frac{(I_{002}/I_{004})}{(I_{002}/I_{004})_{\delta=0}}} \tag{1}$$

where $I_{002}$ and $I_{004}$ are the XRD intensities of (002) sand (004) peak, respectively.

We simulated theoretical PXRD patterns of $ScTaN_2$ structures with varying $\delta$, and calculated the long-range order parameter $S$ for these theoretical structures. The numerical values are reported in Table S1. As displayed in Fig. 3(c), the theoretical $S$ decreases from 1 to 0 as disorder is induced through $\delta$. The experimental order parameter is calculated from the experimental XRD intensity ratio $(I_{002}/I_{004})$ of a film near $x = 0.5$. We found $S_{exp} = 0.87$, which corresponds to $\delta = 0.07$ according to the calculations. Since $\delta = 0.5$ corresponds to complete disorder, this suggests that 7% of the Sc/Ta sites are antisites, that is than 1/10$^{th}$ of all octahedral sites is occupied by Ta atoms, or vice versa. Similar analyses were performed in the case of analogous layered nitride $LiMoN_2$ prepared in bulk form by ammonolysis, where the authors found 15% antisites [31]. Rietveld analysis on various bulk layered oxides such as $LiCoPO_4$ and $LiNiPO_4$ have also led to results in the same range [32,33]. A supplemental analysis is available in the Supplemental Materials file in which we focus on Ta-rich films (see Fig. S6).

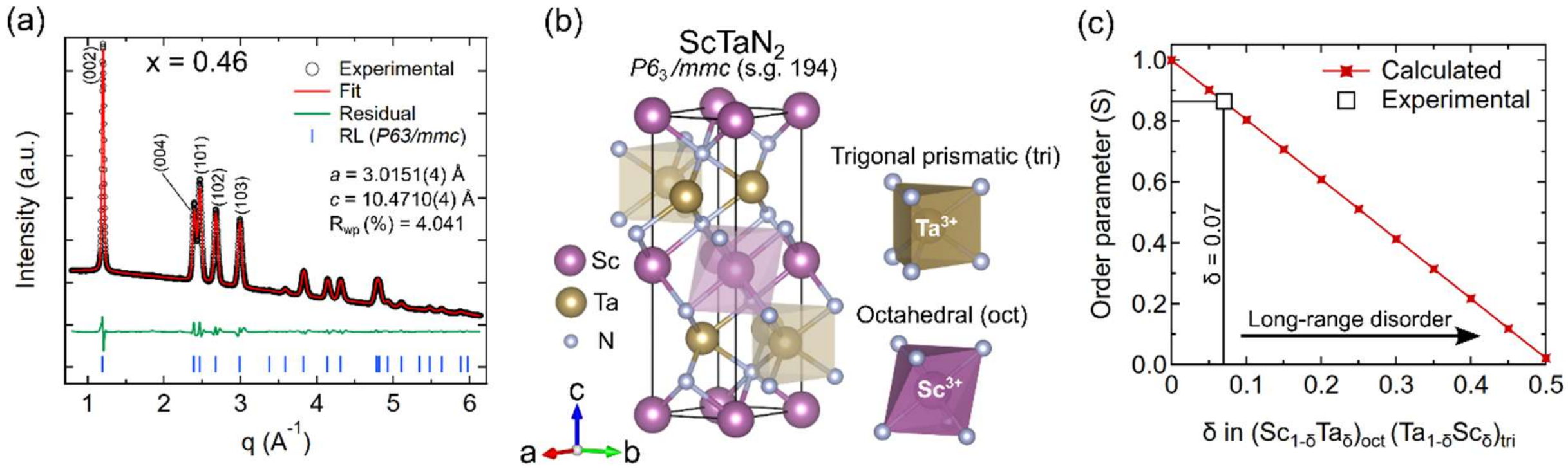

FIG. 3. (a) LeBail refinement of synchrotron GIWAXS pattern acquired near stoichiometric (x = 0.46). The fit is performed in the P63/mmc space group. (b) $ScTaN_2$ RL layered structure (aka "rockseline"). The structure is comprised of alternating sheets of $Sc^{3+}$ in octahedral coordination and $Ta^{3+}$ in trigonal prismatic coordination. (c) Long-range order parameter calculated from the simulated PXRD patterns of $(Sc_{1-\delta}Ta_{\delta})_{oct}(Ta_{1-\delta}Sc_{\delta})_{tri}N_2$ structures. The experimental order parameter is reported on the calculated trace.

### C. Ta-rich $ScTaN_2$ films

At Ta-rich content, the emergence of two additional diffraction peaks at low angle (Fig. 2(b)) revealed the presence of $Ta_5N_6$. To further investigate, we deposited off-stoichiometric $Sc_xTa_{1-x}N$ films, in which $x$ is intentionally shifted to Ta-rich compositions. The as-deposited films with $x$ ranging from 0.12 to 0.28 all exhibit a rocksalt structure, as shown in Fig. S9. The GIWAXS patterns shown in Fig. 4(a) reveal the same characteristic peaks of the RL phase. However, as the film gets enriched with Ta, two additional peaks at $q$ = 1.40 Å$^{-1}$ and $q$ = 1.85 Å$^{-1}$, indicated with an asterisk, gradually increase in intensity. Those can be respectively indexed to the (100) and (102) reflections of $Ta_5N_6$ (space group $P6_3/mcm$, no. 193). $Ta_5N_6$ structure, shown in Fig. 4(c), is quasi-isostructural to $ScTaN_2$ as it exhibits the same layered sequence of octahedra and trigonal prisms. However, each octahedra plane in $Ta_5N_6$ contains one third of Ta vacancies and the in-plane lattice constant is bigger than $ScTaN_2$ ($a$ = 5.18 Å versus 3.06 Å for $ScTaN_2$). Fig. S11 shows the in-plane relations between the two structures. Although the diffraction signature of $Ta_5N_6$ is almost identical to $ScTaN_2$, additional reflections such as (100) and (102) are allowed and can be used as a probe to distinguish the two phases. The presence of $Ta_5N_6$ at low Sc content is further confirmed by the peak analysis presented in Fig. 4(b). As $x$ becomes smaller, the (101), (102), and (103) peaks of $ScTaN_2$ linearly shift to higher $q$, and eventually match with (111), (112), and (113) of quasi-isostructural $Ta_5N_6$. The (002) and (004) peaks also shift linearly (Fig. S8), leading to a contraction of the $c$ parameter by about 1% between $x$ = 0.28 and $x$ = 0.12, therefore agreeing with the bulk value of $Ta_5N_6$ ($c$ = 10.36 Å). We note the rapid drop in intensity of the (002) superlattice peak as the material shifts to $Ta_5N_6$, explained by the loss of contrast within the different layers.

Our GIWAXS results suggest that a smooth structural change occurs as the Sc content is reduced, and from stoichiometric $ScTaN_2$ to $Ta_5N_6$, while maintaining the layered structure. This tie line between the two structures can be explained by supposing that the excess of Ta gradually substitutes Sc in the octahedral sites. The intermediary compounds can be formulated as $(Sc_{1-\alpha}Ta_{\alpha})_{oct}(Ta)_{tri}N_2$ where α represent the excess Ta with respect to stoichiometric.

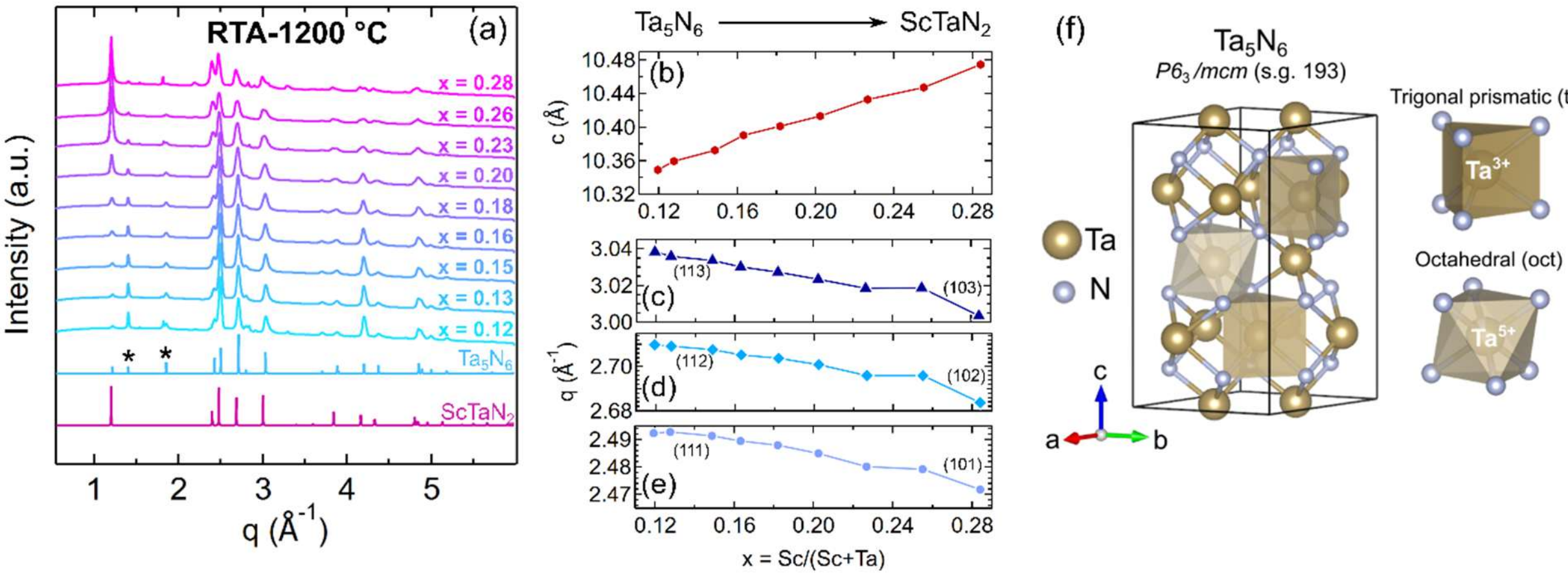

FIG. 4. (a) Synchrotron GIWAXS of non-stoichiometric $Sc_xTa_{1-x}N$ films annealed at 1200 °C for 20 min (λ = 0.97625 Å). (b) Out-of-plane $c$ lattice parameter function of the Sc fraction. Shifting of the diffraction peaks for the fundamental reflections (c) (103), (d) (102) and (e) (103) function of $x$. At low $x$, the peak positions match with (113), (112) and (111) of $Ta_5N_6$. (f) Crystal structure of $Ta_5N_6$, quasi-isostructural to $ScTaN_2$.

Since both Sc and Ta should be in +III oxidation states in $ScTaN_2$, assuming an ionic picture, excess of Ta with respect to Sc should not affect the charge neutrality. Moreover, the ionic radii of $Sc^{3+}$ and $Ta^{3+}$ in a VI coordination environment are relatively equal, facilitating the site substitution of $Sc^{3+}$ by $Ta^{3+}$ in the octahedral sites.

It is interesting to note that such facile Sc/Ta cross-substitution seems to be possible in the Ta-rich region only, as the RL structure is rapidly lost when $x > 0.5$ (see Fig. 2). According to crystal field theory, the effect of a trigonal prismatic coordination on a $d$-metal splits the five degenerate $d$ orbitals into three groups, including the subband $d_{z^2}$ with the lowest energy (Fig. S9). Unlike $Ta^{3+}$, $Sc^{3+}$ does not have the valence electrons ($3d^0$) to fill the $d_{z^2}$ subband and stabilize the trigonal prismatic coordination, therefore an octahedral coordination is preferred [34,35]. This could explain why excess $Sc^{3+}$ fails to stabilize the layered structure and instead favors a disordered rocksalt structure.

Several studies have demonstrated the stabilization of non-stoichiometric $ABN_2$ layered nitrides, enriched with 4d or 5d metals (such as W, Ta, Nb, and Mo). Table 1 presents some of those materials. It includes, for example, $Fe_{0.8}W_{1.2}N_2$, $Fe_{0.8}Mo_{1.2}N_2$, $Co_{0.6}Mo_{1.4}N_2$, and $Li_{0.84}W_{1.16}N_2$ [36–39]. One study synthesized Fe-deficient $Fe_{0.72}WN_2$ by leaching Fe from a $FeWN_2$ precursor. Interestingly, unlike the stoichiometric compound, they measured room-temperature ferromagnetism in the Fe-deficient sample, with a small saturation magnetization but a high coercivity (~0.8 T) [40,41]. One reason could be the partial suppression of the geometric spin frustration in the Fe triangular lattice (octahedral layer) caused by Fe vacancies, revealing an interesting pathway for designing new ferromagnetic materials. These results reveal the ability of layered $ABN_2$ nitrides to accommodate non-stoichiometry without amorphization or phase separation. It also makes $ABN_2$ compounds good potential candidates for (de)intercalation chemistry [42]. The nitrides discussed above all contain early 4d or 5d transition metals on the *B* site, such as Nb, Mo, Ta, or W. By the large number of electrons in their *d* orbitals, those metals have strong covalent bonding character and can accommodate different oxidation states in transition metal nitrides, oxides, sulfides, phosphides or carbides. Therefore, they can form various phases and adapt in different coordination environments. One could explain why Nb, Mo, Ta, or W can easily relocate in the octahedral sites and stabilize *B*-rich layered nitrides in the $ABN_2$ form. Furthermore, by their high atomic number, heavy elements such as Ta (Z = 73) or W (Z = 74) can contribute to spin-orbit coupling, likely inducing exotic electronic properties. This has been demonstrated in $MgTa_2N_3$, which becomes a Dirac semimetal when spin-orbit coupling in considered [43,44]. Finally, it is interesting to point out that for each 4d/5d metal discussed above, there is a corresponding binary nitride $B_5N_6$ featuring the same layered structure as $Ta_5N_6$ discussed in this work. An additional discussion is available in the Supplemental Materials file.

TABLE I. Examples of reported *A*-*B*-N ternary nitrides featuring a "rockseline" structure (or equivalent). The compounds are classified by their transition metal (*TM)* on the *B* site. Non-stoichiometric compounds are also reported to highlight their ability to stabilize in the *B*-rich region.

| *TM* on *B* site | Material | Space group | Ref. |
|---|---|---|---|
| Ta | $ScTaN_2$ | $P6_3/mmc$ | [23,24], this work |
| | $MgTa_2N_3$ | $P6_3/mcm$ | [42,45] |
| | $LiTa_3N_4$ | $P6_3/mcm$ | [45] |
| | $Ta_5N_6$ | $P6_3/mcm$ | [46] |
| Mo | $MgMoN_2$ | $P6_3/mmc$ | [13,47–49] |
| | $CoMoN_2$ | $P6_3/mmc$ | [38] |
| | $Co_{0.6}Mo_{1.4}N_2$ | $P6_3/mmc$ | [38] |
| | $FeMoN_2$ | $P6_3/mmc$ | [50,51] |
| | $Fe_{0.8}Mo_{1.2}N_2$ | $P6_3/mmc$ | [37] |
| | $MnMoN_2$ | $P6_3/mmc$ | [52] |
| | $LiMoN_2$ | $P6_3/mmc$ | [31] |
| | $Mo_5N_6$ | $P6_3/mcm$ | [53,54] |
| Nb | $ScNbN_2$ | $P6_3/mmc$ | [24] |
| | $MnNb_2N_3$ | $P6_3/mcm$ | [55] |
| | $LiNb_2N_3$ | $P6_3/mcm$ | [56] |
| | $Nb_5N_6$ | $P6_3/mcm$ | [46,55] |
| W | $FeWN_2$ | $P6_3/mmc$ | [52] |
| | $Fe_{0.8}W_{1.2}N_2$ | $P6_3/mmc$ | [40,41] |
| | $LiWN_2$ | $P6_3/mmc$ | [57] |
| | $Li_{0.84}W_{1.16}N_2$ | $P6_3/mmc$ | [57] |
| | $MgWN_2$ | $P6_3/mmc$ | [14] |
| | $W_5N_6$ | $P6_3/mcm$ | [58] |

### D. Film morphology of $ScTaN_2$ film

The film morphology of stoichiometric $ScTaN_2$ was characterized before and after annealing by electron microscopy. Imaging was performed at locations in the combinatorial films near stoichiometric. Fig. 5(a) and Fig. 5(c) show the cross-section SEM micrographs of as-deposited films and films annealed by RTA at 1200 °C. The as-deposited film exhibits columnar grains, typically observed in a sputtered film. The film thickness is measured at 227 ± 3 nm and the average width of columnar grains at 15 ± 5 nm. The sputtered $SiN_x$ capping layer is clearly visible on the as-deposited film, with a thickness of 23 ± 3 nm, as well as the $SiN_x$ insulating layer present on the Si substrate.

In Fig. 5(c), the cross-section micrograph of the annealed film shows that the columnar grain morphology is lost, coalescence of the grains has occurred, and the interfaces are roughened. Although darker regions indicate some small presence of voids, the film remains dense, and no significant cracks or fractures are observed. Small delamination from the substrate is visible and is likely due to thermal mismatch stress induced by the extreme thermal processing and more specifically the fast cooling. The thickness of the $ScTaN_2$ film is measured around 195 ± 6 nm, which is about 32 nm thinner than the as-deposited film. This could be a consequence of grains densification as well as recrystallisation. Surprisingly, the $SiN_x$ thin capping layer is no longer visible, suggesting it vaporized during RTA. EDS maps of the as-deposited film (Fig. 5(b)) show uniform distribution of Sc, Ta and N throughout the film. The capping layer nature is verified by the detection of some Si on the surface. The presence of oxygen is detected mostly near the surface across the capping layer, which is probably due to post-deposition air exposure. This agrees with the AES depth profile, displayed in Fig. 5(d), showing the anionic ratio O/(O+N) across the same film and revealing a low concentration of oxygen in the film, with O/(O+N) ≈ 2 %. However, a significant amount of oxygen was detected at the sample surface, but it rapidly dropped across the $SiN_x$ capping layer, confirming that oxygen from the surface has diffused in through the capping layer. This demonstrates the importance and effectiveness of using barrier layers for nitride films to minimize contamination from air exposition.

### E. Electrical and magneto-transport

Electrical and magneto-transport measurements were performed on a $ScTaN_2$ annealed film near stoichiometry, which exhibits the layered phase. The temperature-dependent longitudinal resistivity ($\rho_{xx}$) presented in Fig. 6a. Interestingly, we observe two main regimes: i) from 300 K to 100 K, the resistivity decreases, similar to a metallic-type conduction mechanism dominated by phonon scattering, and approaches a crossover point at 100 K; ii) below 100 K, the resistivity follows an upturn and start increasing while cooling down. This crossover suggests that an additional scattering mechanism kicks in at low temperatures. None of the conventional conduction mechanism (such as Arrhenius or variable range hopping) can be fitted in this region. Therefore, this upturn could be due to a quantum correction of the conduction due to weak anti-localization (WAL) as discussed later. The 300 K magnitude of $\rho_{xx}$ of 0.51 mΩ.cm paired with the small change with temperature ($\rho_{xx}(300K)/\rho_{xx}(100K)$ = 1.08) could indicate either a narrow bandgap semiconducting or a semimetallic behavior. This observation agrees with reported density of states and bandgap calculations on $ScTaN_2$ [23,59]. Furthermore, the dependance of $\rho_{xx}$ in the 100 – 300 K regime is relatively similar to what was measured in isostructural $FeWN_2$ epitaxial films [50].

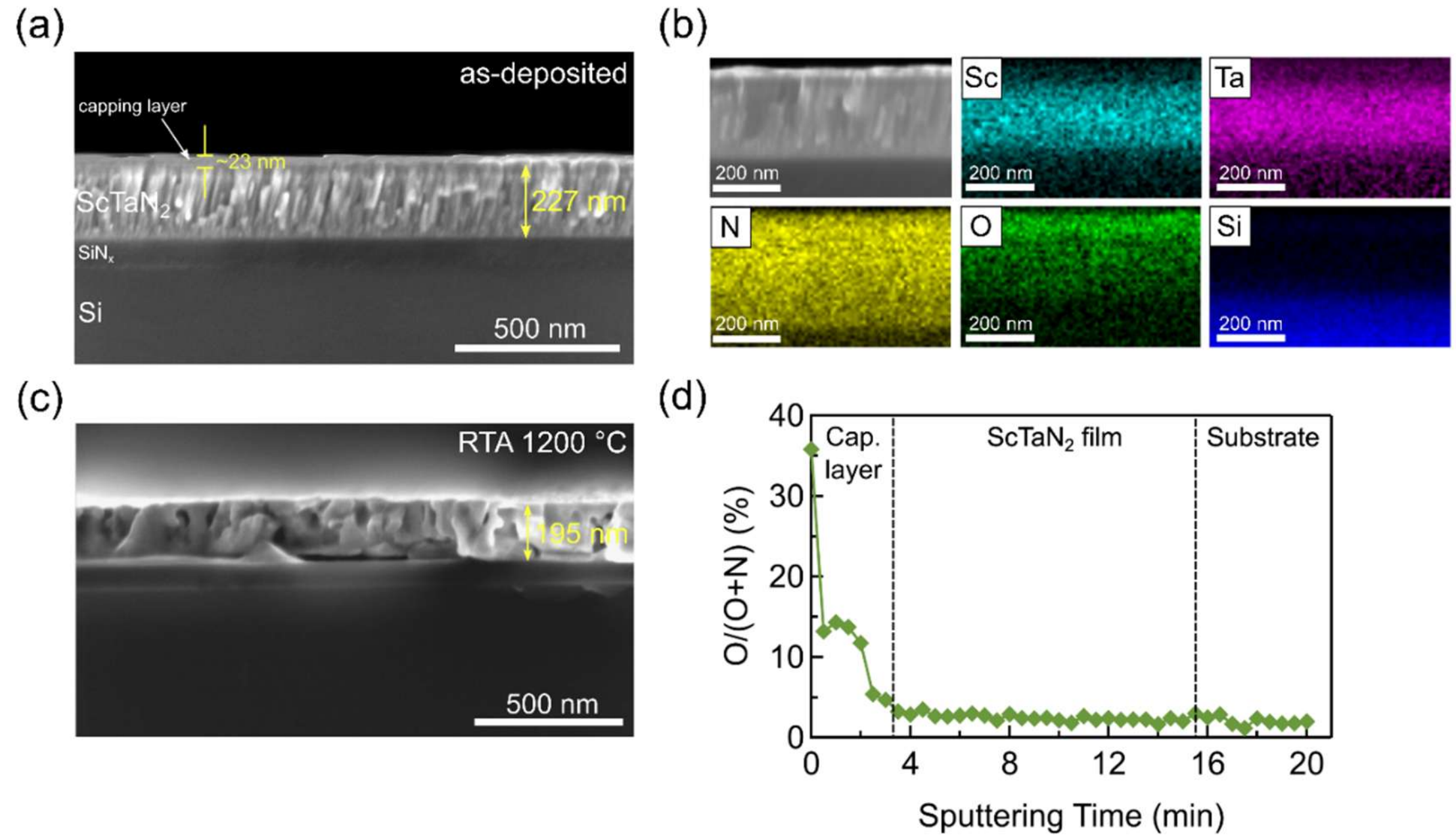

FIG. 5. SEM cross-section images of (a) as-deposited and (c) annealed $ScTaN_2$ films. (b) Cross-sectional SEM-EDS maps of a stoichiometric as-deposited $ScTaN_2$ film. (d) AES depth profile showing the relative anionic concentration O/(O+N) throughout the as-deposited film, taken at near-stoichiometry condition. SEM-EDS maps of the annealed film are shown in Fig. S2.

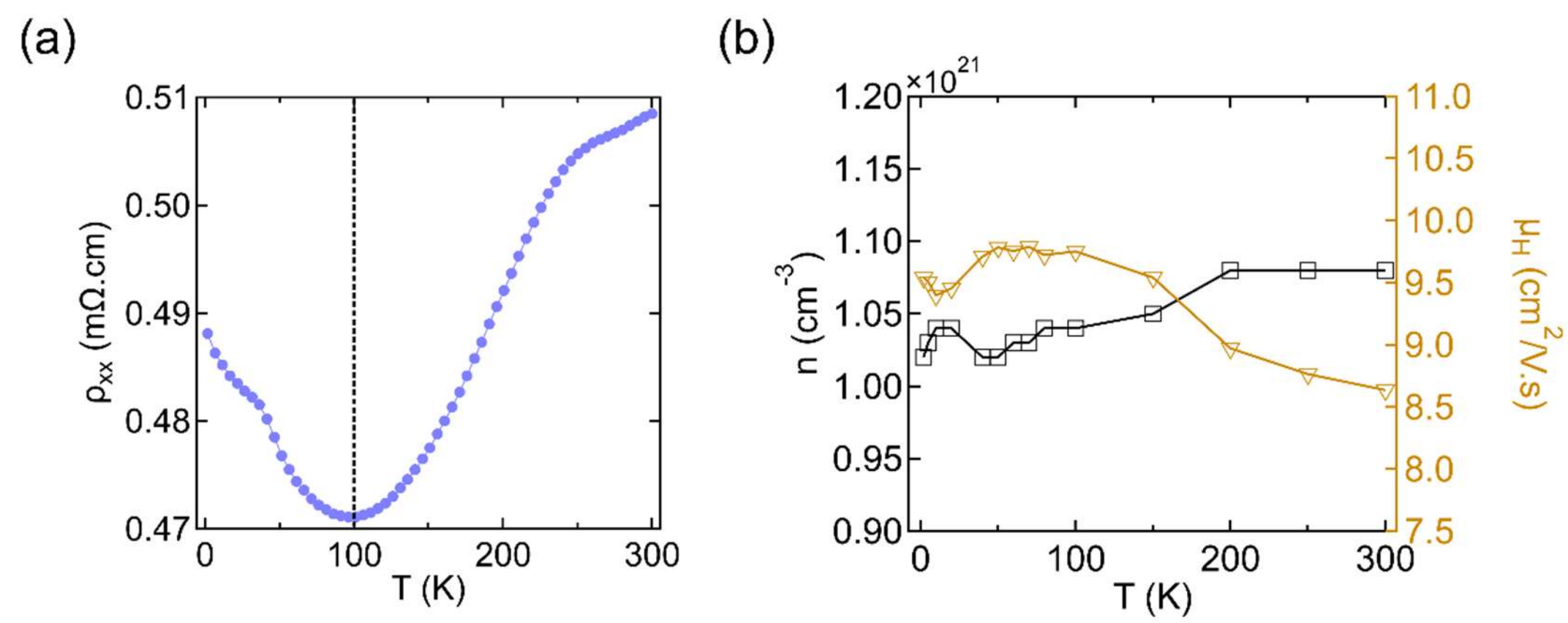

FIG. 6. Electrical and magneto-transport characterization of $ScTaN_2$ film. (a) Temperature dependance of the longitudinal electrical resistivity. The vertical dash line indicates the conduction regime turnover. (b) Hole-like carrier concentration and carrier mobility obtained from Hall effect.

From Hall resistivity measurements (Fig. 6b), we find a nearly-temperature-independent, hole-like carrier density $n$ around $10^{21}$ $cm^{-3}$, and both are nearly temperature independent. The Hall mobility $\mu_H$ is 9.5 $cm^2V^{-1}s^{-1}$ at 2 K and slightly increases to reach a plateau between 50 K and 100 K, consistent with the crossover observed in $\rho_{xx}$, and then decreases as temperature increases, due to scattering. The low mobility observed likely originates from the presence of grain boundaries due to the polycrystalline nature of the $ScTaN_2$ film, as well as defects. Additionally, the in-plane Seebeck coefficient was measured at room-temperature and found to be positive with a low value of +19 μV/K, which confirms the hole-type conduction. The small in-plane Seebeck coefficient can be attributed to the high carrier density and the low mobility, and agrees well with previous calculations [59].

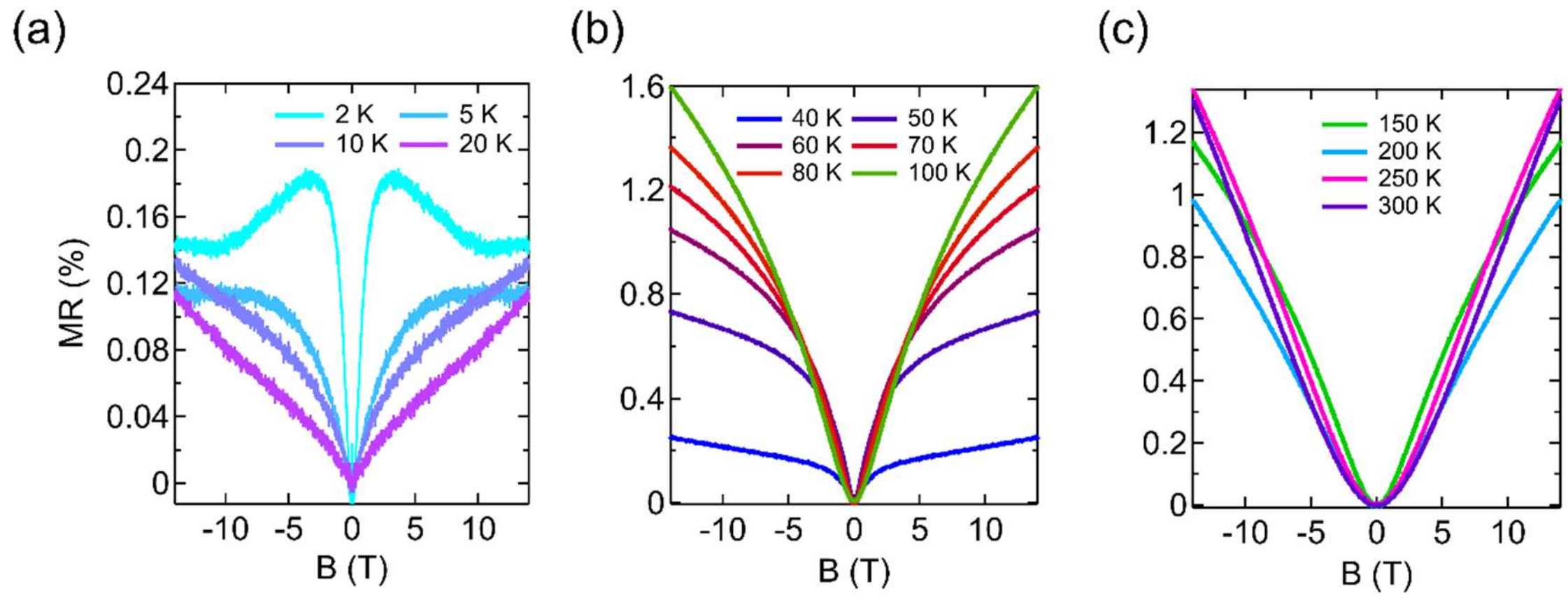

FIG. 7. Magnetoresistance (MR) as a function of the perpendicular magnetic field in regions of temperatures of (a) 2-20 K, (b) 40-100 K and (c) 150-300 K.

The magnetoresistance ($MR\ (\%) = [R(H) - R(0)]/R(0)\ \times\ 100$) measured at different temperatures is shown in Fig. 7. For T > 150 K, the MR shows a quadratic dependance at low field but then exhibits a crossover beyond which the MR is positive linear and non-saturating, as suggested with fitting analysis shown in Fig. S10. As temperature decreases below 150 K, the MR reveals a mixed behavior with a logarithmic (or sublinear) dependance at low field and a linear dependance at high field. The distinction between the two regimes becomes more pronounced as the temperature decreases, noted by the emergence of a cusp-like feature at low field which is often attributed to WAL, a quantum effect leading to destructive interference of the spin-dependent electron wavefunctions, usually induced by strong spin-orbit coupling. This effect is best captured at 2 K, where the MR sharply increases at low field. At high field, the non-saturating linear magnetoresistance (LMR) regime could likely arise from strong disorder, where the magnetoresistance is controlled by fluctuation of carrier mobilities, as described by Parish and Littlewood classical model [60,61]. Interestingly, this disorder-induced LMR effect has been observed in polycrystalline materials with narrow or zero bandgap [62,63]. However, we point out that the magnitude of MR is small and only reaches a maximum of 1.6% at 100 K, likely because of the low mobility. The MR could also be significantly reduced from scattering events due to the high carrier density ($10^{21}$ $cm^{-3}$), thus limiting the field-induced cyclotron motion of carriers ($\tau_{scat} \ll \tau_{cycl}$). This explanation agrees with the polycrystalline nature of $ScTaN_2$ films in this work. Furthermore, this is a sputtered film, annealed under extreme conditions, and likely has a defective microstructure.

The low mobility in our $ScTaN_2$ films likely originates from the presence of structural defects and grain boundaries. The high carrier density observed results in the semi metallicity of $ScTaN_2$, implying that the Fermi level lies within the overlapping conduction and valence bands. The presence of defects such as antisites is likely to move the Fermi level deeper, creating more states, and thus increasing further the carrier density Previous theoretical studies on $ScTaN_2$ have shown that the main states around the top of the valence originate from the presence Ta 5d orbitals, whereas the bottom of the conduction band is mostly comprised of a mix of Sc 3d and Ta 5d orbitals [23,59]. Thus, off-stoichiometry or, more specifically, Ta substitution would make hole conduction more likely. In addition, the presence of Ta in a higher valence state than $Ta^{3+}$, such as $Ta^{5+}$ (Fig. 4f) which is suggested by the average valence state of 3.6 for Ta in $Ta_5N_6$, could bring extra positive charge to the system. Moreover, it is known that WAL is directly linked to spin-orbit coupling in materials. In the case of $ScTaN_2$, a strong spin-orbit coupling could be expected, likely to originate from the Ta 5d orbitals. This explains why WAL is observed at low temperature. It would not be surprising for $ScTaN_2$ to exhibit similar topological properties as $MgTa_2N_3$, mentioned previously [43,44]. The structural similarities as well as spin orbit coupling would likely yield to similar electronic states, and even possibly Dirac nodes in the band structure. These questions will be investigated in future work. For instance, growing high-quality epitaxial $ScTaN_2$ films of on matching substrates, such as hexagonal SiC ($a$ = 3.07 Å) or AlN ($a$ = 3.11 Å) will be necessary to improve electronic properties and confirm these hypotheses.

## IV. CONCLUSION

We successfully synthesized compositionally graded Sc-Ta-N thin films using combinatorial RF sputtering and demonstrated the conversion of disordered rocksalt structure into the cation-ordered $ScTaN_2$ layered structure upon rapid thermal annealing at 1200 $^{o}$C. The results revealed that phase transformation occurs at the $ScTaN_2$ stoichiometry and in the Ta-rich composition region, accompanied by a strong (002) superlattice peak confirming the long-range ordering. We estimated the long-range order parameter $S$ to be 0.86 for stoichiometric $ScTaN_2$. This study also highlighted the ability of $ScTaN_2$ to accommodate large off-stoichiometry in the Ta-rich region, facilitated by the formation of quasi-isostructural $Ta_5N_6$ at lower Sc content and stabilizing the layered structure over a wide composition range. Magneto-transport characterizations reveal $ScTaN_2$ to be a semimetal or a narrow bandgap semiconductor, agreeing with previous calculations, whereas the conduction is limited by defects and impurities in our polycrystalline samples. The observation of weak anti-localization at low temperature suggests that strong spin-orbit coupling occurs in $ScTaN_2$, likely originating from Ta. This could potentially lead to exotic topological properties such as in Dirac semimetal $MgTa_2N_3$. This work provides insights into the role of disorder and stoichiometry in layered nitrides deposited as thin films and reveals $ScTaN_2$ as a promising candidate for exploring exotic electronic states and correlated transport phenomena.

## ACKNOWLEDGMENTS

This work was authored in part at the National Renewable Energy Laboratory (NREL) for the U.S. Department of Energy (DOE) under Contract No. DE-AC36-08GO28308. Funding was provided by the Office of Science (SC), Basic Energy Sciences (BES), Materials Chemistry program, as a part of the Early Career Award "Kinetic Synthesis of Metastable Nitrides" (materials growth, synchrotron and most characterization/analysis); and by DOE-SC-BES, Division of Materials Science, through the Office of Science Funding Opportunity Announcement (FOA) Number DE- FOA-0002676: Chemical and Materials Sciences to Advance Clean-Energy Technologies and Transform Manufacturing (electronic transport measurements/analysis). Use of the Stanford Synchrotron Radiation Lightsource, SLAC National Accelerator Laboratory is supported by DOE's SC, BES under Contract No. DE-AC02-76SF00515. The views expressed in the article do not necessarily represent the views of the DOE or the U.S. Government.

Supplemental Materials

# Structural stability, elemental ordering, and transport properties of layered $ScTaN_2$

Baptiste Julien, Ian A. Leahy, Rebecca W. Smaha, John S. Mangum, Craig L. Perkins, Sage R. Bauers, Andriy Zakutayev

*Materials Science Center, National Renewable Energy Laboratory, 15013 Denver West Parkway, Golden, Colorado 80401, United States.*

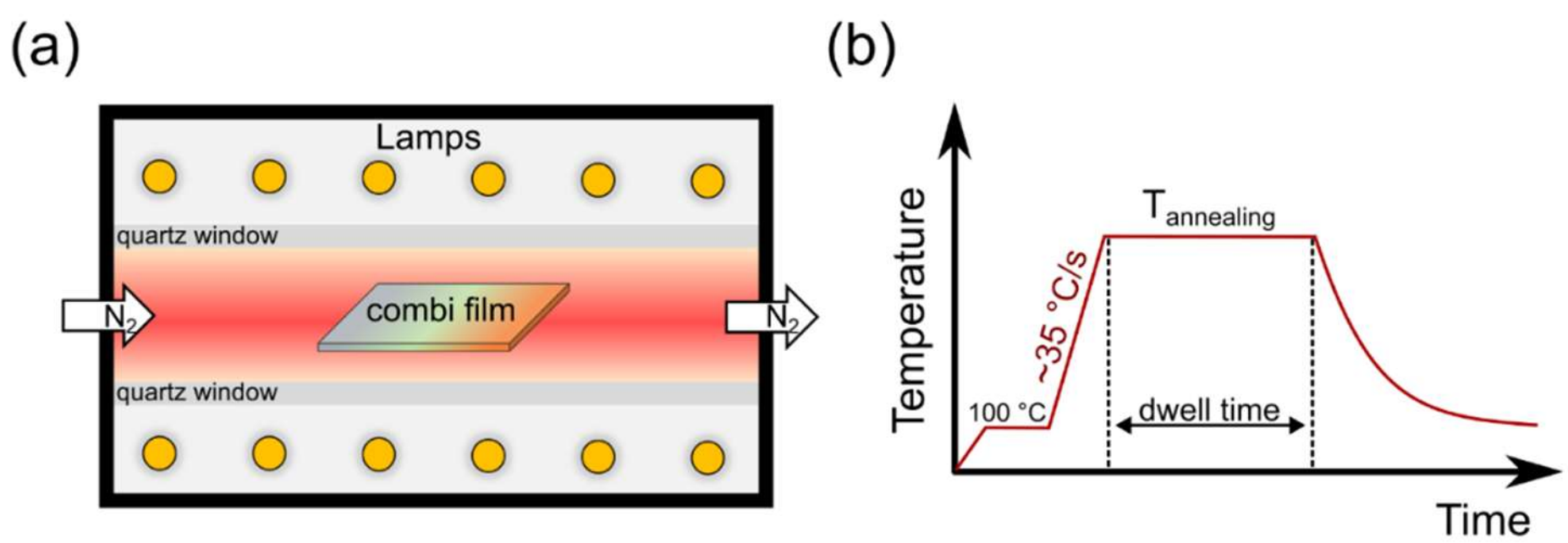

**Figure S1**: (a) Schematic of the RTA furnace used for annealing combinatorial films. (b) Time-dependent temperature profile during an RTA run.

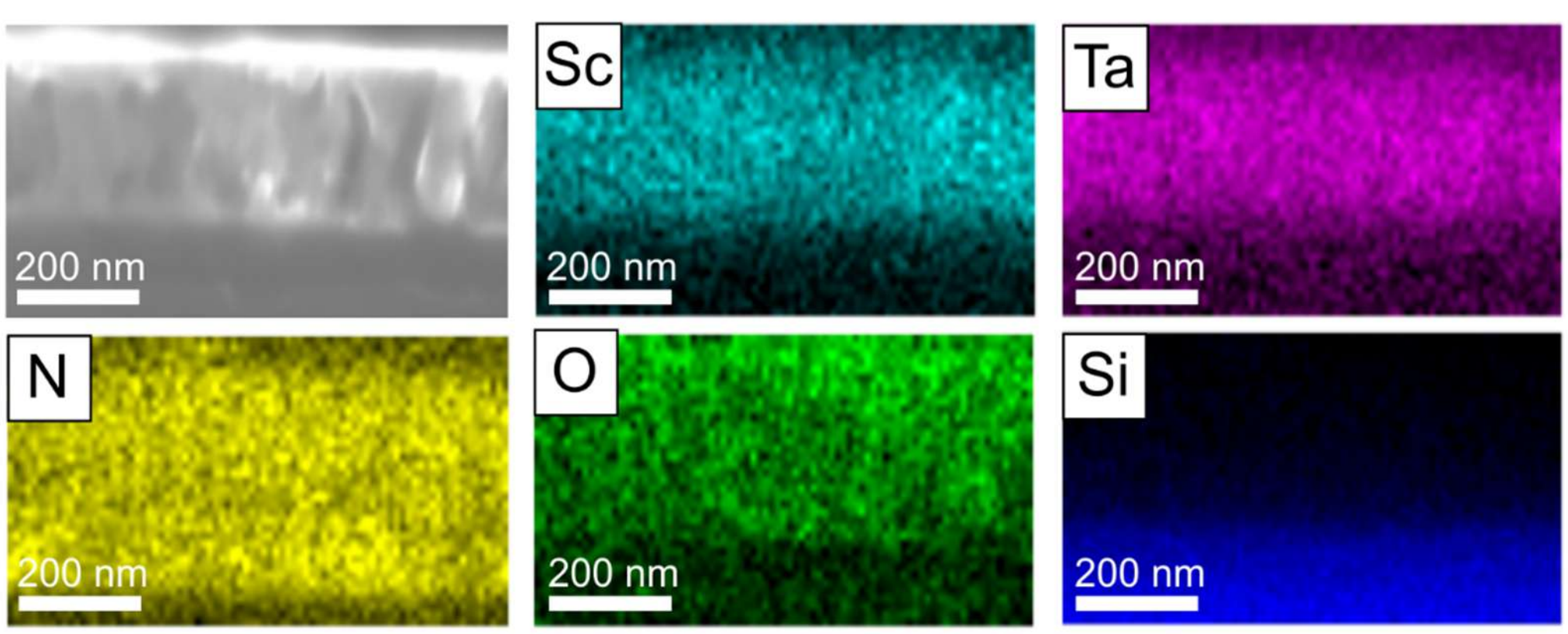

**Figure S2**: Cross-sectional SEM-EDS maps of a $ScTaN_2$ film after RTA at 1200 °C.

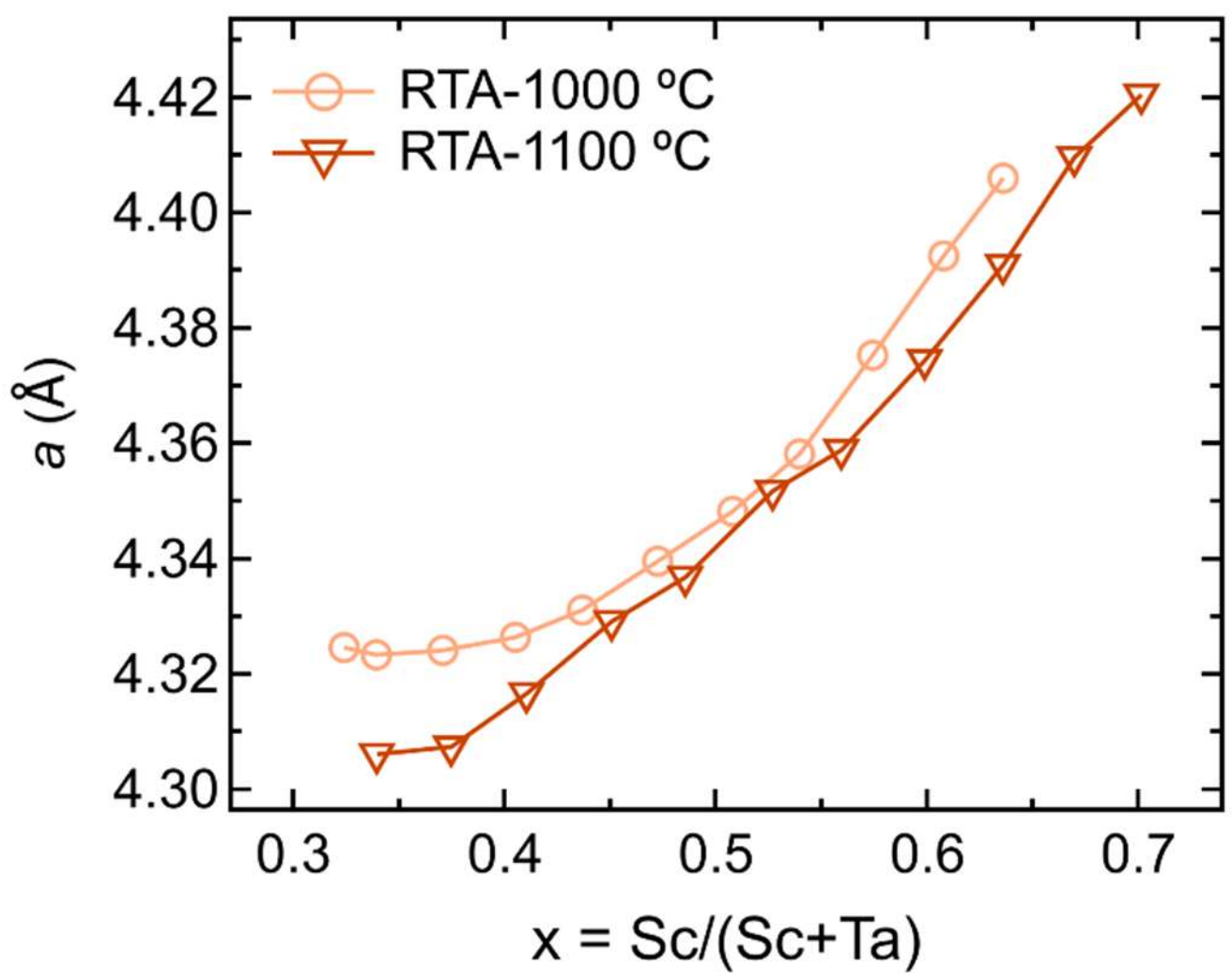

**Figure S3**: Lattice parameter of RS-$ScTaN_2$ at 1000 °C and 1100 °C annealing temperatures.

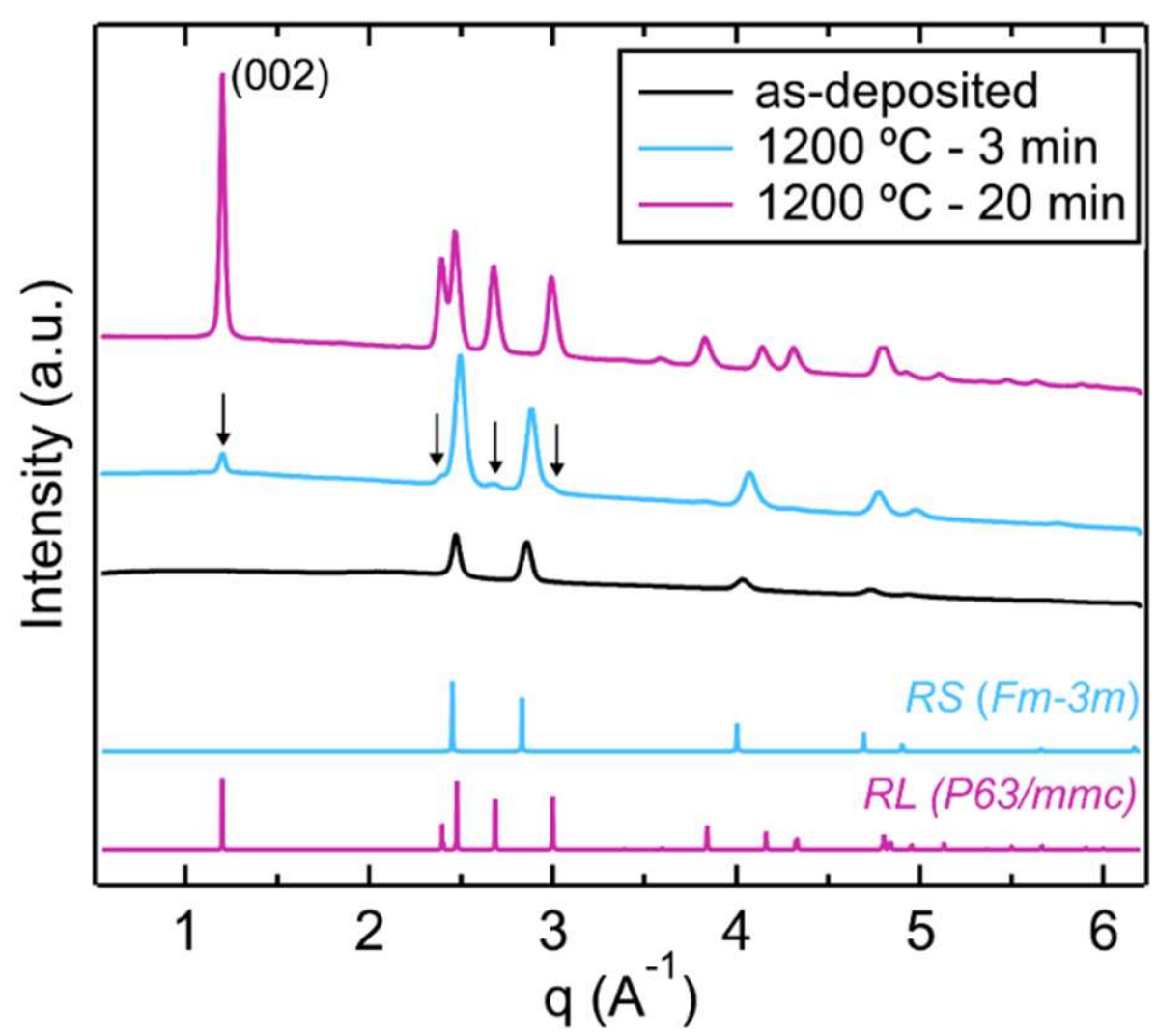

**Figure S4**: Synchrotron GIWAXS of $Sc_xTa_{1-x}N$ films with $x \approx 0.5$, after RTA for 3 min and 20 min, showing the importance of annealing time in the transformation of the RS phase. Arrows point at low-intensity RL peaks, showing that RL starts to form in the film after 3 min annealing. However, the conversion is not completed, and the film remains mostly in the RS structure.

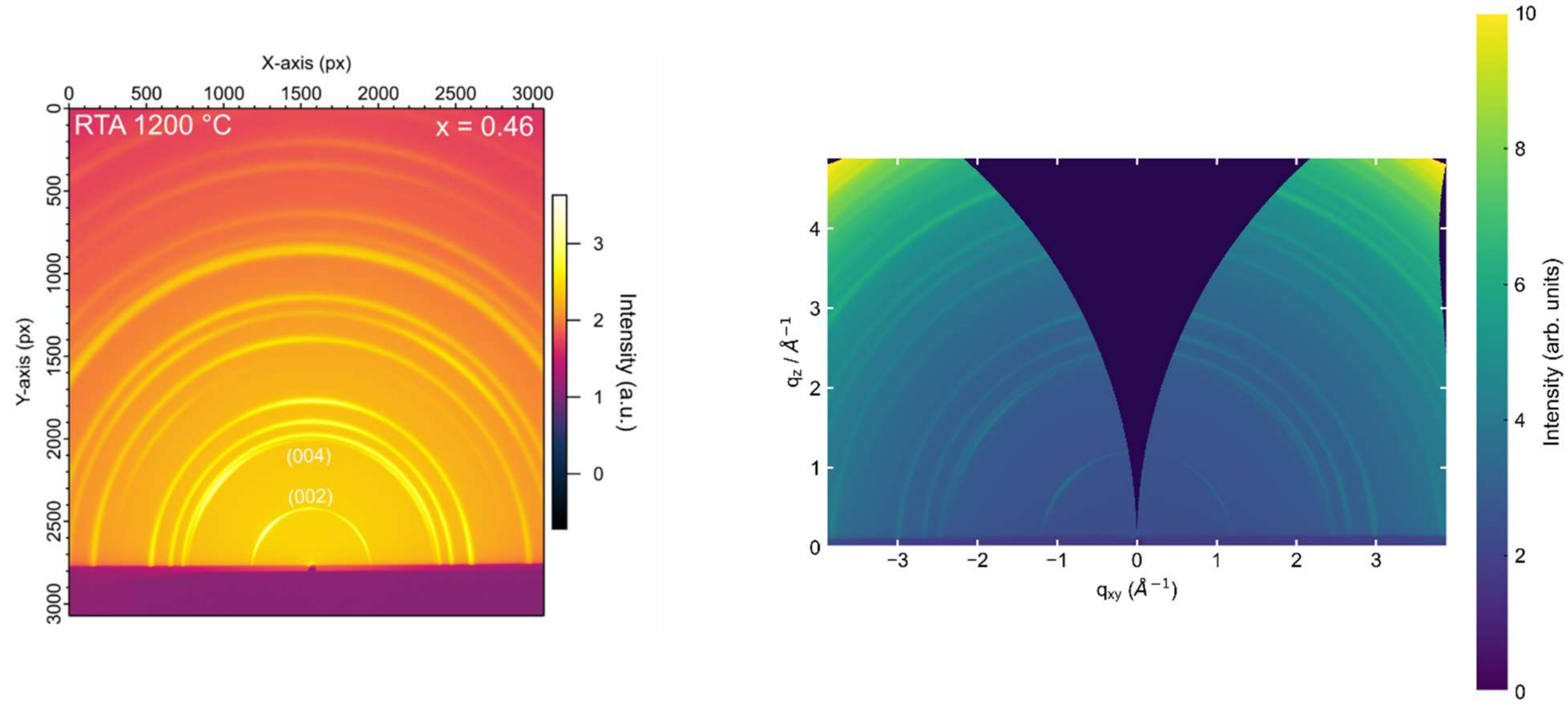

**Figure S5**: Synchrotron GIWAXS raw detector image (left) and in reciprocal space after wedge correction (right) of a near-stoichiometry ($x$ = 0.46) $ScTaN_2$ film after RTA at 1200 °C for 20 min. The diffraction image shows characteristic Debye-Scherrer rings of the layered $ScTaN_2$ phase. The superstructure (002) ring as well as (004) are indicated on the image. A certain degree of texture is present as indicated by concentric rings. In $q$-space, the missing wedge cuts off diffraction intensity in the out of plane direction, leading to a misleading when

**Table S1**. Calculated (00*l*) reflection intensities for different fraction of antisites δ in $(Sc_{1-\delta}Ta_{\delta})_{oct}(Ta_{1-\delta}Sc_{\delta})_{tri}$ for stoichiometric $ScTaN_2$ ($x$ = 0.5). The long-range parameter $S$ is extracted from Eq. (1) in the main text. The bottom line shows the experimental peak intensities measured from the GIWAXS data for a film at $x$ = 0.5, and the corresponding order parameter. Simulations were performed using the Python library *pymatgen*, with λ = 0.97625 Å. For these calculations, the isotropic thermal parameter B was set to 1 for each atom.

| **Fraction of antisites** | **(00*l*) calculated intensities and long-range order parameter** | | | **Site occupancy** | | | | |
|---|---|---|---|---|---|---|---|---|
| **δ** | $I_{002}$ | $I_{004}$ | $S$ | oct-Sc | oct-Ta | tri-Ta | tri-Sc | N |
| 0 | 242.7 | 100 | 1 | 1 | 0 | 1 | 0 | 1 |
| 0.05 | 197.5 | 100 | 0.902 | 0.95 | 0.05 | 0.95 | 0.05 | 1 |
| 0.1 | 157.0 | 100 | 0.804 | 0.9 | 0.1 | 0.9 | 0.1 | 1 |
| 0.15 | 121.1 | 100 | 0.706 | 0.85 | 0.15 | 0.85 | 0.15 | 1 |
| 0.2 | 89.9 | 100 | 0.609 | 0.8 | 0.2 | 0.8 | 0.2 | 1 |
| 0.25 | 63.3 | 100 | 0.511 | 0.75 | 0.25 | 0.75 | 0.25 | 1 |
| 0.3 | 41.4 | 100 | 0.413 | 0.7 | 0.3 | 0.7 | 0.3 | 1 |
| 0.35 | 24.1 | 100 | 0.315 | 0.65 | 0.35 | 0.65 | 0.35 | 1 |
| 0.4 | 11.4 | 100 | 0.217 | 0.6 | 0.4 | 0.6 | 0.4 | 1 |
| 0.45 | 3.5 | 100 | 0.119 | 0.55 | 0.45 | 0.55 | 0.45 | 1 |
| 0.5 | 0.11 | 100 | 0.021 | 0.5 | 0.5 | 0.5 | 0.5 | 1 |
| **Experiment** | **181.82** | **100** | **0.865** | **0.93** | **0.07** | **0.93** | **0.07** | **1** |

***Discussion – Stabilization of the RL phase for Ta-rich films***

Analysis of diffraction results have shown that, the $ScTaN_2$ RL structure is maintained in a wide range of composition in the Ta-rich region of $Sc_xTa_{1-x}N$ ($0.35 \leq x \leq 0.50$). On the other hand, when the Sc fraction is slightly higher than stoichiometric, the RL phase is rapidly lost, and the Sc-rich film therefore remains in a RS structure. This suggests that the formation of layered motifs upon annealing is highly sensitive to excess of Sc but can easily accommodate excess of Ta. One could hypothesize that in the case of Ta-rich stoichiometry, the excess Ta would substitute a fraction of Sc in the octahedral sites, inducing short-range disorder but maintaining the long-range ordered RL structure, to some extent. To investigate this hypothesis, we simulated new RL structures, this time varying the stoichiometry $x$ in the Ta-rich region ($x < 0.5$). The excess Ta is inserted in the octahedral sites along with Sc, and thus an average octahedral site is now partially occupied by Sc and Ta. Therefore, we can define Ta-rich $Sc_xTa_{1-x}N$ structures for $0 \leq x \leq 0.5$ as:

$$Sc_xTa_{1-x}N \Leftrightarrow (Sc_{2x}Ta_{1-2x})_{oct}(Ta_1Sc_0)_{tri}N_2$$

We note that here, no antisite defects are considered, so the trigonal prismatic layer is assumed to be perfectly filled with Ta ($O_{tri\text{-}Ta} = 1$, $O_{tri\text{-}Sc} = 0$), and only the octahedral layer is mixed with Sc and Ta, following the overall stoichiometry. The structures are assumed to be free of vacancies, thus the total occupancy of a site cannot be lower than 1. For example, $x = 0.35$ corresponds to a $Sc_{0.7}Ta_{1.3}N_2$ stoichiometry where 30% of the octahedral sites are occupied with Ta ($O_{oct\text{-}Sc} = 0.7$ and $O_{oct\text{-}Ta} = 0.3$), as illustrated in Fig S7. In the simulated patterns of $Sc_{0.7}Ta_{1.3}N_2$, the (002) superstructure peak intensity decreases by about 60% for $x = 0.35$ with respect to the stoichiometric case, whereas the (004) peak intensity is almost not affected. The $I_{002}/I_{004}$ ratio calculated from simulated patterns is plotted as a function of $x$ and shown in Fig. S5(b) and compared with the experimental data extracted from the GIWAXS data. As expected, the $I_{002}/I_{004}$ ratio from simulated structures rapidly drops as Ta substitutes Sc in the octahedral layer. At stoichiometric $ScTaN_2$ ($x = 0.5$), the difference between the data and the simulation can be explained in term of antisite defects inducing short-range disorder and therefore negatively affecting (002) intensity, as explained in the main text. However, the parabolic-like trend observed in the simulations does not match with the trend observed in the diffraction data. Interestingly, the (002)/(004) ratio exhibits a saturation-like trend between $x = 0.42$ and $x = 0.50$, which does not agree with our simple model in which only Ta substitutes Sc in the octahedral sites. The discrepancy between the experimental data and the simulation clearly indicates that this simple model of Sc substitution by Ta in $Sc_xTa_{1-x}N$ cannot fully explain the structural transition observed.

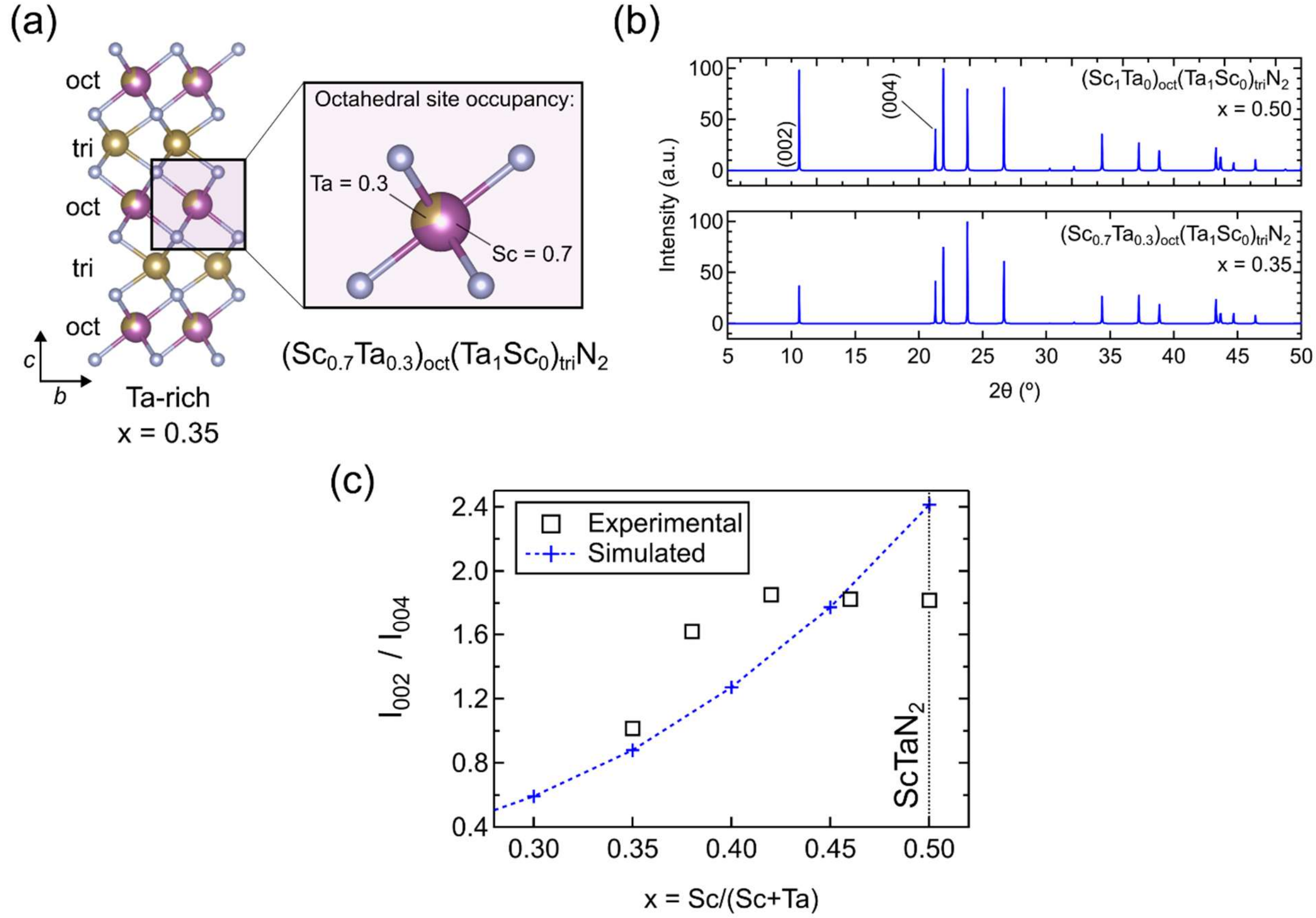

**Figure S6** (a) Two-dimensional representation of $(Sc_{0.7}Ta_{0.3})_{oct}(Ta_1Sc_0)_{tri}N_2$ corresponding to a stoichiometry of $Sc_{0.7}Ta_{1.3}N_2$ ($x$ = 0.35). The inset illustrates the mixed occupancy in the octahedral sites where a fraction of Sc is substituted by the excess Ta. (b) Simulated PXRD patterns of $ScTaN_2$ and $Sc_{0.7}Ta_{1.3}N_2$. (c) Peak intensity ratio $I_{002}/I_{004}$ plotted versus the stoichiometry $x$ in the Ta-rich region, from the simulated PXRD patterns as well as the experimental data.

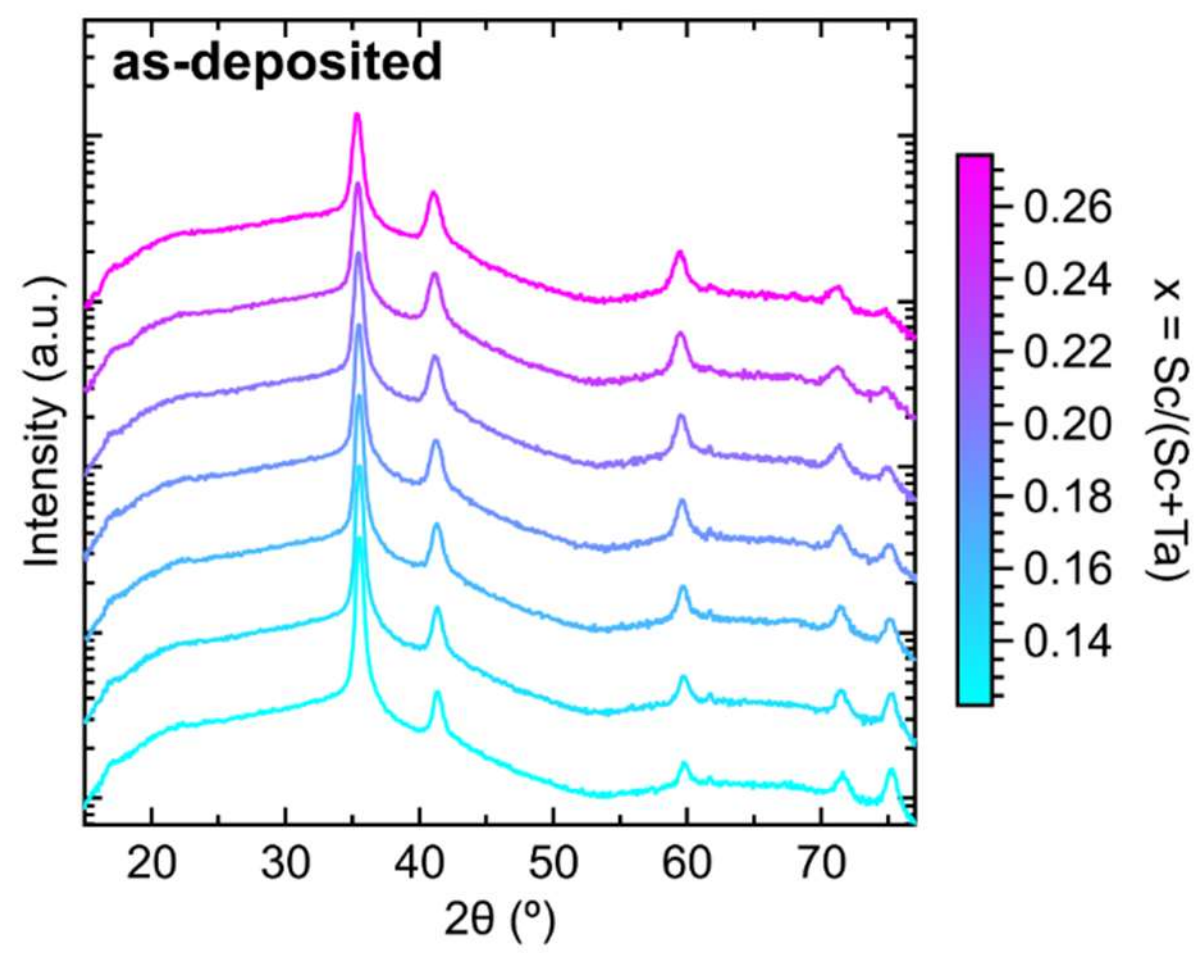

**Figure S7**: Laboratory XRD patterns of as-deposited Ta-rich $Sc_xTa_{1-x}N$ films, revealing a rocksalt structure for all compositions.

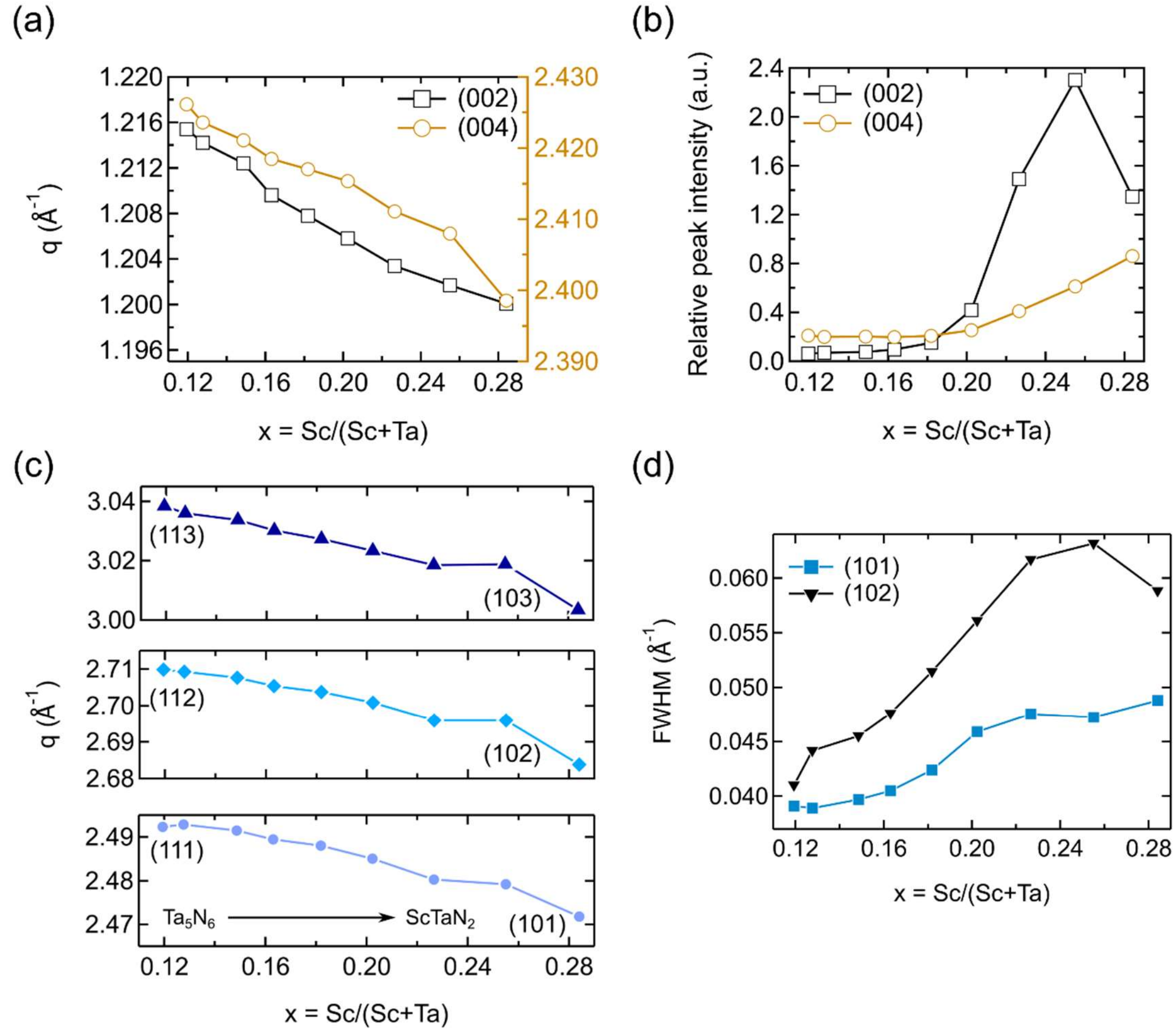

**Figure S8**: Additional peak analysis extracted from the GIWAXS data of Ta-rich $Sc_xTa_{1-x}N$ films with RL structure. (a) Peak position and (b) peak intensity of the (002) and (004) reflections. (c) Peak position of the main reflections, shifting from $Ta_5N_6$ (111)-like to $ScTaN_2$ (101)-like. (d) Full width half maximum (FWHM) of the (101) and (102) peaks.

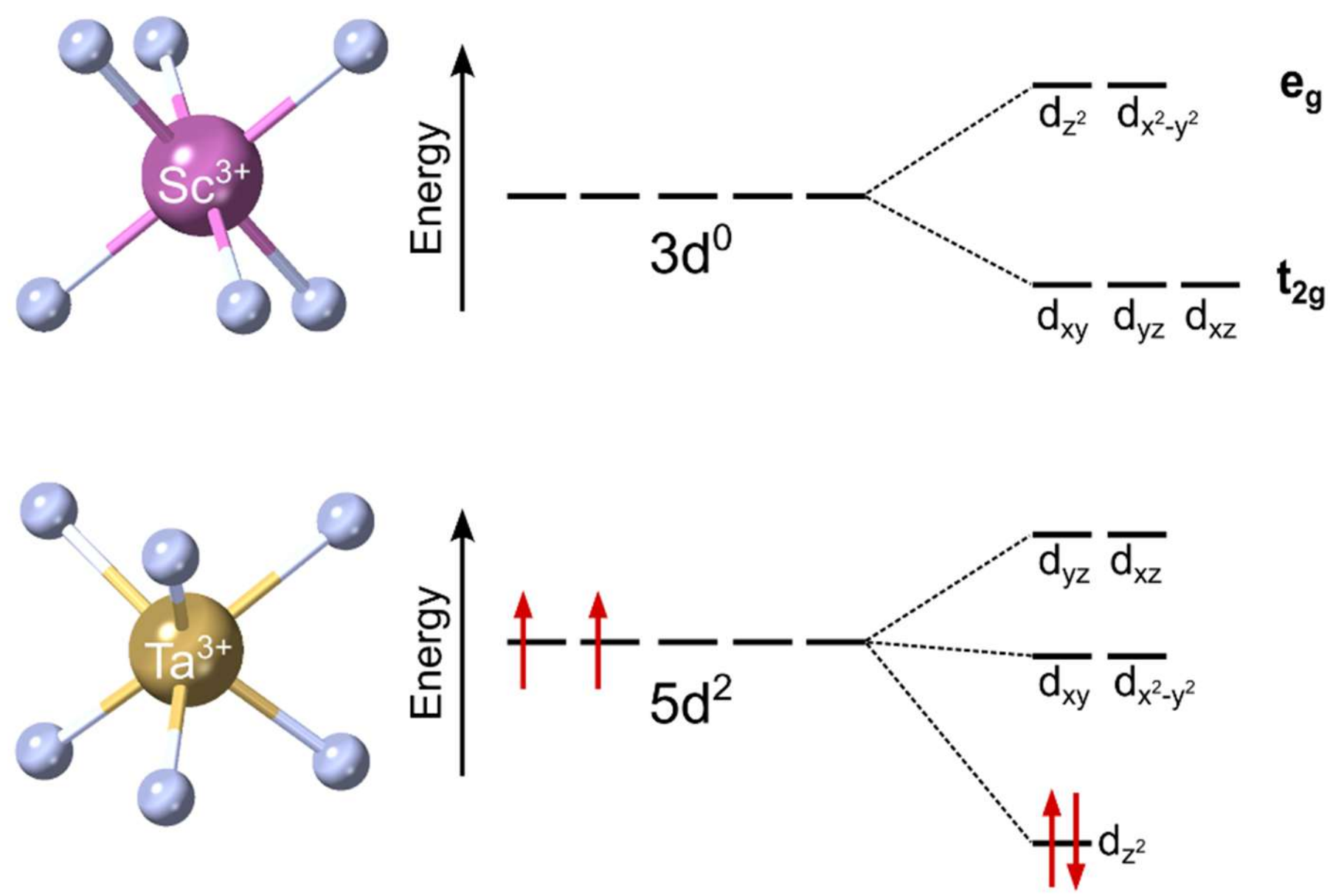

**Figure S9**: Crystal field splitting of *d* orbital energy levels of (a) octahedrally coordinated $Sc^{3+}$ and (b) a trigonal prismatic coordinated $Ta^{3+}$.

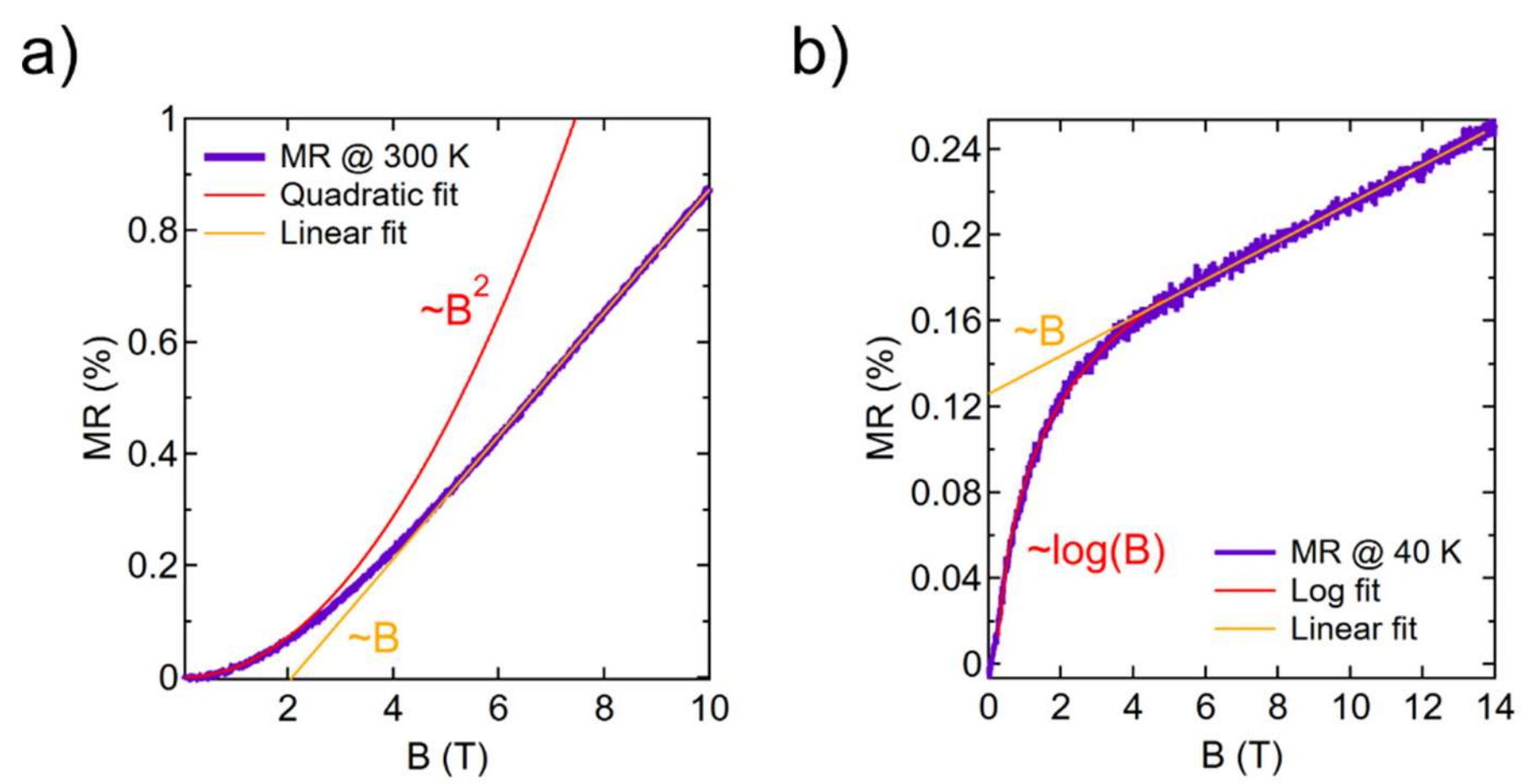

**Figure S10**: (a) Magnetoresistance at 300 K, showing a quadratic dependence at low fields and a linear tendency at higher fields. (b) Magnetoresistance at 40 K, showing a logarithmic dependance at low fields, indicative of WAL, and a linear tendency at higher fields.

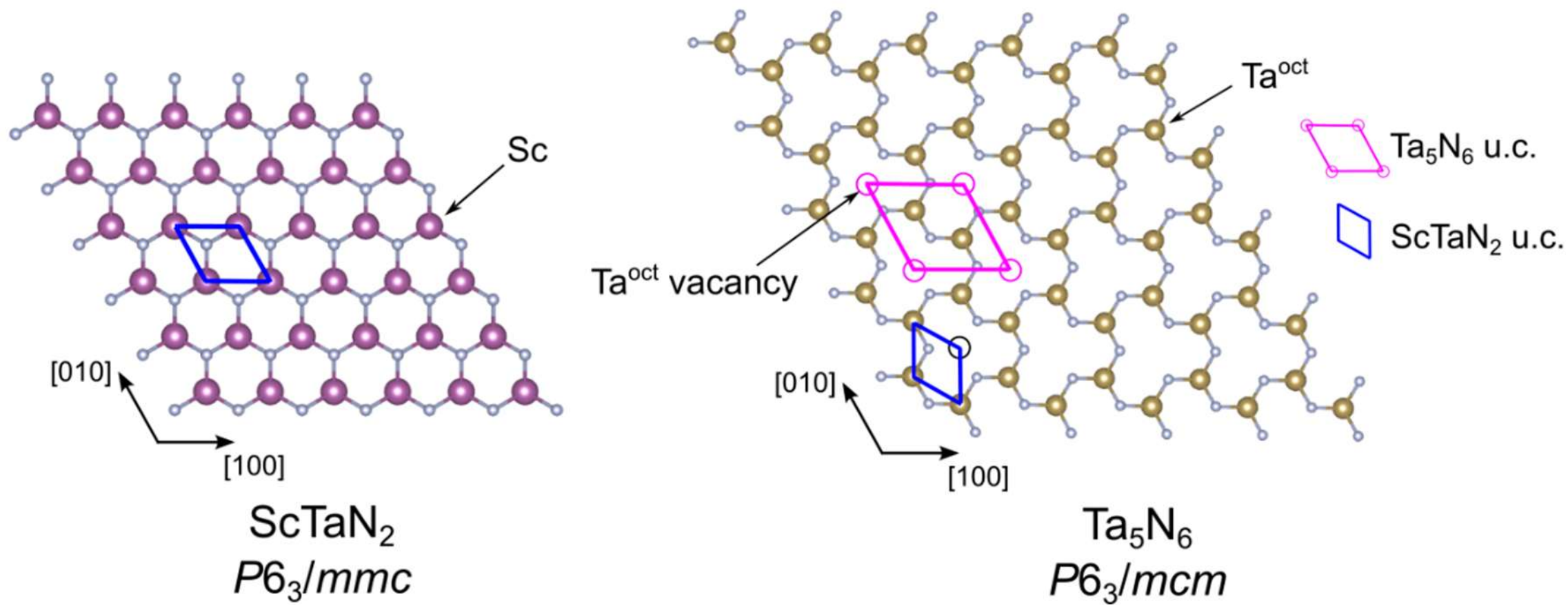

**Figure S11**: In-plane relation between $ScTaN_2$ and $Ta_5N_6$ structures. In $ScTaN_2$, the (001) plane is composed of Sc sitting in each octahedral sites whereas in the case of $Ta_5N_6$, one third of the Ta octahedral sites is vacant. As shown in the figure on the right, the unit cell of $ScTaN_2$ can be fitted within the $Ta_5N_6$ lattice with a 30° in-plane rotation.

### ***Additional discussion on the interplay between $ScTaN_2$ and $Ta_5N_6$***

Among numerous reported phases in the Ta-N system, $Ta_5N_6$ is the most stable nitrogen-rich phase, with a formation energy of -1.339 eV/atom. $Ta_5N_6$ belongs to $P6_3/mcm$ hexagonal space group and features a layered structure with $a$ = 5.184 Å and $c$ = 10.366 Å. It is isostructural to $MgTa_2N_3$. It is quasi-isostructural to $ScTaN_2$ and consists of alternating layers of octahedral and trigonal prismatic coordinated Ta, with Ta vacancies located in one third of the octahedral sites [1–3]. We can therefore refer $Ta_5N_6$ as a RL-like structure and we can formulate it as $(Ta_{0.66}\,\square_{0.33})_{oct}(Ta)_{tri}N_2$ where □ denotes a vacancy. A similar relation exists for isostructural Ta-rich nitrides such as $MgTa_2N_3$ or $LiTa_3N_4$ which can be related to $(Mg_{0.67}Ta_{0.33})_{oct}(Ta)_{tri}N_2$ and $(Li_{0.5}Ta_{0.5})_{oct}(Ta)_{tri}N_2$ respectively [4,5], and can be referred as RL 1-2-3 or RL 1-3-4 structures. In those materials, Ta occupies the trigonal prismatic layer plus a fraction of the octahedral sites along with the alkali metal. This is somewhat similar to Ta-rich $Sc_xTa_{1-x}N$ in which the excess Ta substitute a fraction of the octahedral Sc. Interestingly, in the case of highly Ta-rich films ($x$ < 0.20), we could imagine that the residual Sc is located on the Ta vacancy sites □ in $Ta_5N_6$. In fact, $Ta_5N_6$ can be written as $\square_{0.33}\,Ta_{1.66}\,N_2$ which is reminiscent of non-stoichiometric $Sc_xTa_{1-x}N$ where the vacancy □ could be replaced with Sc. For example, $x$ = 0.15 stoichiometry corresponds to $Sc_{0.3}Ta_{1.7}N_2$ and can then be related to $Ta_5N_6$. Furthermore, to maintain charge neutrality, the ionic picture of $Ta_5N_6$ is described as $(Ta^{5+}_{0.66}\,\square^{-}_{0.33})_{oct}(Ta^{3+})_{tri}\,N^{3-}_{2}$. Inserting Sc in place of the vacancy □, and assuming Sc to be trivalent ($Sc^{3+}$), would require lowering the oxidation state of oct-Ta from $Ta^{5+}$ to $Ta^{3+}$. The ability of Ta to change its oxidation state would facilitate this process, explaining the smooth transition from $Ta_5N_6$ to $ScTaN_2$. This could explain the tie line of phase transformation we observed between $ScTaN_2$ and $Ta_5N_6$ when the Sc fraction is significantly reduced.

It is interesting to note that for Ta, Mo, and Nb based nitrides in the $AB$N$_2$ form, there is a corresponding binary $B_5N_6$ with the same ground-state structure as some $AB_2N_3$ nitrides, such as $MgTa_2N_3$ or $MnNb_2N_3$.

Thus, a tie line of stabilization might exist as the material becomes *B*-rich such as $ABN_2 \rightarrow AB_2N_3 \rightarrow B_5N_6$, and a smooth structural transformation occurs.